\documentclass[preprint2]{aastex}

\def\eps{\epsilon}
\newcommand{\be}{\begin{equation}}
\newcommand{\ee}{\end{equation}} 

\shorttitle{Cosmic-ray Diffusion}
\shortauthors{Fatuzzo et al.}

\begin{document}

\title{High Energy Cosmic-ray Diffusion in Molecular Clouds: \\ A Numerical Approach}

\author{M. Fatuzzo}
\affil{Physics Department, Xavier University, Cincinnati, OH 45207}
\email{fatuzzo@xavier.edu}

\author{F. Melia}
\affil{Department of Physics, The Applied Math Program, and Steward Observatory, \\ 
The University of Arizona, AZ 85721}
\email{melia@physics.arizona.edu}

\author{E. Todd}
\affil{Physics Department, The University of Arizona, AZ 85721}
\email{etodd@physics.arizona.edu}

\and

\author{F. C. Adams}
\affil{Michigan Center for Theoretical Physics, University of Michigan \\
Physics Department, Ann Arbor, MI 48109}
\email{fca@umich.edu}

\begin{abstract}
The propagation of high-energy cosmic rays through giant molecular clouds constitutes a
fundamental process in astronomy and astrophysics. The diffusion of cosmic-rays through 
these magnetically turbulent environments is often studied through the use of energy-dependent 
diffusion coefficients, although these are not always well motivated theoretically. Now,
however, it is feasible to perform detailed numerical simulations of the diffusion process
computationally. While the general problem depends upon both the field structure and particle 
energy, the analysis may be greatly simplified by dimensionless analysis. That is, for a 
specified purely turbulent field, the analysis depends almost exclusively on a single parameter
 -- the ratio of the maximum wavelength of the turbulent field cells to the particle 
gyration radius. For turbulent magnetic fluctuations superimposed over an underlying uniform 
magnetic field, particle diffusion depends on a second dimensionless parameter that 
characterizes the ratio of the turbulent to uniform magnetic field energy densities. 
We consider both of these possibilities and parametrize our results to provide simple
quantitative expressions that suitably characterize the diffusion process within 
molecular cloud environments. Doing so, we find that the simple scaling laws often 
invoked by the high-energy astrophysics community to model cosmic-ray diffusion through 
such regions appear to be fairly robust for the case of a
uniform magnetic field with a strong turbulent 
component, but are only valid up to $\sim 50$ TeV particle energies for
a purely turbulent field.  These results have 
important consequences for the analysis of cosmic-ray processes based on 
TeV emission spectra associated with dense molecular
clouds.
\end{abstract}


\section{Introduction}

Observations of $\gamma$-rays associated with regions of dense molecular gas 
provide important clues about how cosmic-rays (CR's) are injected within our galaxy.    
However, a proper treatment of this problem requires an understanding of how
CR's diffuse through turbulent environments.  While this subject
has received considerable attention since the 
pioneering works of Jokipii (1966) and Kulsrud \& Pearce (1969),
the exact nature of particle transport remains unresolved. 

A standard approach to the problem invokes the use of 
the spherically symmetric diffusion equation
\begin{equation}
\frac{\partial f}{\partial t} = \frac{D}{R^2} \frac{\partial}{\partial R} R^2 \frac{\partial f}{\partial R} +
  \frac{\partial}{\partial E_p} \left( Pf \right) + Q \;,
\end{equation}
where $f \equiv f(E_p, R, t)$ is the distribution of particles as a function of energy, distance, and time; $P = 
-(dE_p/dt)$ is the continuous energy loss rate; $Q \equiv Q(E_p, R, t)$ is the source function; and 
$D \equiv 
D(E_p)$ is the energy-dependent diffusion coefficient. 
A simplified solution to this diffusion equation may be obtained by assuming a power-law injection spectrum, 
$f_{\rm inj} \propto E_p^{-\alpha}$, and a power-law diffusion coefficient,
\be
D(E_p) = D_{10} \left({E_p\over 10 \,{\rm GeV}}\right)^{\delta}\,,
\ee
in the 
energy regime where $\tau_{pp}$ is independent of energy
(we note that values of $\delta = 1/2$ and $D_{10}$ $\sim$ $10^{26-28}$
cm$^2$ s$^{-1}$ are typically assumed for molecular cloud
environments---see, e.g., Aharonian \& Atoyan 1996; Torres 
et al. 2003; Gabici et al. 2009). As shown by Aharonian and Atoyan (1996), the 
solution to the diffusion equation in such a case can be approximated as:
\begin{eqnarray}
f\left( E_p, R, t \right) \approx \qquad\qquad\qquad\nonumber \\
\frac{N_0 E_p^{-\alpha}}{\pi^{3/2} {R_{\rm diff}}^3} \exp\left( -\frac{\left( \alpha 
- 1 \right)t}{\tau_{pp}} - \frac{R^2}{{R_{\rm diff}}^2} \right) \;,
\end{eqnarray}
where
\begin{eqnarray}
R_{\rm diff} \equiv R_{\rm diff}(E_p,t) = 
\qquad\qquad\nonumber\\
2 \sqrt{D(E_p)\,t \frac{\exp \left(t\delta / \tau_{pp}\right) - 1}{t\delta / 
\tau_{pp}}}
\end{eqnarray}
is the ``diffusion radius" corresponding to the radius of the sphere out to which particles with energy $E_p$ 
effectively propagate after a time $t$.  In the limit that $t \ll \tau_{pp}$, the ``diffusion radius'' simplifies to 
$R_{\rm diff} = 2\sqrt{D(E_p) \, t}$.

In this paper, we investigate how high-energy CR's propagate through molecular 
cloud-like environments by instead using a modified numerically based formalism 
developed for the general study of cosmic-ray diffusion by Giacalone \& Jokipii (1994).
This formalism has already been used to study the transport of cosmic rays
in chaotic magnetic fields with Kolmogorov turbulence (Casse et al. 2002) 
and has been applied successfully in several specific contexts (see,
e.g., Kowalenko \& Melia 1999; Casse et al. 2002; De Marco et al. 2007; 
Wommer et al. 2008; Fraschetti \& Melia 2008). 

The first goal of this work is to extend the general treatment of Casse et al. (2002)
by exploring a greater dynamic range of wavelengths over which turbulence acts
and by considering Kraichnan, Bohm and Kolomogorov turbulence for two magnetic field 
configurations:  
1) a purely turbulent field; and 2) a uniform magnetic field with a strong turbulent 
component.  
The second goal of this work is to provide a baseline analysis for 
the propagation of $\sim 1 - 10^4$ TeV cosmic-rays in molecular cloud environments.

As we shall see, CR diffusion in purely turbulent fields depends primarily on a single 
dimensionless parameter
\be
\bar \lambda_{{\rm max}} \equiv {\lambda_{{\rm max}}\over R_g}\,,
\ee
where 
$\lambda_{\rm max}$ represents the longest turbulent field wavelength and $R_g$ is the 
particle gyration radius in a uniform field of the same magnetic energy density as 
that of the turbulent field.   This parameter is related to the particle rigidity $\rho$ through 
the expression $\bar\lambda_{{\rm max}} = 2\pi/\rho$.
In the second case, CR diffusion also depends on a second
dimensionless parameter---the ratio of turbulent field energy density to the uniform 
field energy density. As we shall see, the result of our work indicates that the 
diffusion coefficients often invoked to describe CR diffusion through molecular cloud 
environments appear to be valid for $\la 50$ TeV cosmic rays propagating in
a purely turbulent field, and appear to be fairly robust for the case of a
uniform magnetic field with a strong turbulent 
component.  

Our paper is organized as follows. The relevant properties of molecular clouds are 
briefly reviewed in \S 2, where we also outline our treatment of these environments. 
The scheme for generating the turbulent magnetic field is presented in \S 3, and the 
equations that govern the motion of CR's are dimensionalized in \S 4. Solutions to 
these equations are presented in \S 5 for purely turbulent fields, 
and in \S 6 for a uniform field with a strong 
turbulent component. We compare and contrast the results of our work to those of Casse et al. (2002)
in \S 7.  We then consider what effects our results have on previous treatments 
of CR diffusion through molecular clouds in \S 8, and summarize our work in \S 9.

\section{Giant Molecular Cloud Environments}
Typical giant molecular clouds (GMCs) contain a total mass of $\sim$$10^5\;M_\odot$
within physical size scales of tens of parsecs, and, as such, have mean densities
of $n_{H_2}$$\sim$$100$ cm$^{-3}$.  However, these large complexes are highly nonuniform, 
exhibiting hierarchical structure that can be characterized in terms of clumps 
($R$$\sim$$1$ pc, $n_{H_2}$$\sim$$10^3$ cm$^{-3}$) and dense cores ($R$$\sim$$0.1$ 
pc, $n_{H_2}$$\sim$$10^4$--$10^5$ cm$^{-3}$) surrounded by an interclump gas of 
density $n_{H_2} \sim 5$--$25$ cm$^{-3}$.

Exactly how the magnetic field is partitioned within GMCs is not yet known. In the simplest 
case, where flux freezing applies, the magnetic field strength $B$ in the interstellar
medium would scale with the gas density $n_{H_2}$ according to $B\propto n_{H_2}^{1/2}$.
It is noteworthy, then, that an analysis of magnetic field strengths measured 
in molecular clouds yields a relation between $B$ and $n_{H_2}$ of the form
\begin{equation}
B \sim 10 \,\mu\hbox{\rm G} \left({n_{H_2} \over 10^2\, \hbox{\rm cm}^{-3}}
\right)^{0.47}\;,
\end{equation}
though with a significant amount of scatter in the data used to produce this fit 
(Crutcher 1999; but see also Basu 2000). This result is consistent with the idea 
that nonthermal linewidths, measured to be $\sim$$1$ km s$^{-1}$ throughout the 
cloud environment (e.g., Lada et al. 1991), arise from MHD fluctuations.  

The exact nature of the magnetic turbulence is not well-constrained, although magnetic 
fluctuations are typically assumed to have a power-law spectrum such that their intensity 
at a given wavenumber scales according to $(\delta B_k)^2 \sim k^{-\Gamma}$, 
with indices typically taken to be $\Gamma = 1$ (Bohm), $\Gamma = 3/2$ (Kraichnan) 
or $\Gamma = 5/3$ (Kolmogorov). In addition, the range in wavelengths over which 
these fluctuations occurs is not well known, although it is reasonable to assume that
the upper end corresponds to the lengthscale over which the fluctuations are
generated. (For example, in the ISM, the turbulence is generated by supernova
remnants and stellar-wind collisions, so one might expect the longest wavelength 
to be on the order of several parsecs or less.) Also, the lower end probably corresponds 
to the scale at which the magnetic field couples most effectively to the particles, 
i.e., on the order of several gyration radii, since this is where the magnetic field 
loses most of its energy. 

Given the complexities and uncertainties in the global properties of the magnetic 
field structure within GMCs, we make several simplifying assumptions throughout 
this baseline work. Specifically, we assume a homogeneous medium and that all MHD 
fluctuations propagate with a uniform (Alfv\'enic) speed $v_A =$ 1 km s$^{-1}$. 
Although much of our analysis is dimensionless and therefore easily scaled, we adopt
fiducial values when dimensionalizing our results. Specifically, we assume that
magnetic fluctuations have a maximum wavelength of $\lambda_{\rm max}$ = 1 pc (essentially
the typical distance between stellar wind sources, as noted above). Further, we consider 
both the case of a purely turbulent field and the case of an underlying uniform magnetic 
field with a strong turbulent component. For the former case, we assume that the energy 
density of the turbulent field is equal to that of a 10 $\mu$G uniform field. For 
the latter, we assume that the underlying uniform field has a magnetic strength 
of $B_0$ = 10 $\mu$G, and that the turbulent component has the same energy density as 
the uniform field.  
 
\section{The Turbulent Magnetic Field}

A novel numerical method for analyzing the fundamental physics of ionic motion
in a static turbulent magnetic field was presented by Giacalone \& Jokipii (1994),
who showed that ions in complete 3D situations readily cross the resulting magnetic 
field. We generalize this pioneering work by considering time-dependent fluctuations 
that propagate with a uniform speed $v_A$ (as first attempted in a different context 
by Fraschetti \& Melia 2008). Within this framework, the magnetic field through 
which cosmic rays of mass $m$ and charge $q$ propagate is expressed in terms of 
the gyration frequency via the parameter ${\bf \Omega} ({\bf r}, t)  = q 
{\bf B}({\bf r},t)/mc$. The total field is then written as the sum of a static 
background component ${\bf \Omega_b}({\bf r})$ and a fluctuating, time dependent 
component $\delta{\bf \Omega}({\bf r},t)$, but we note that it is not necessary 
to have a background component, and for cases where such a component exists, 
fluctuations need not be small. Further, a time-dependent turbulent electric field 
$\delta {\bf E}({\bf r},t)$ must also be present (as required by Faraday's law;
Fraschetti \& Melia 2008). As shown below, $\delta E << \delta B$ for molecular 
cloud environments and, as such, the effects of such an electric field may be 
ignored in the analysis presented here.

The turbulent magnetic field is generated by summing over a large number $N$ of 
randomly polarized transverse waves of wavelength $\lambda_n = 2\pi / k_n$:
\begin{eqnarray}
\delta {\bf \Omega}({\bf r},t) = \sum_{n = 1}^{N} \Omega_n 
\left[\hbox{\rm cos}\, \alpha_n
\hat y' \pm i \,\hbox{sin}\, \alpha_n \hat z' \right] \nonumber \\
\,\hbox{exp}\left[i k_n (x'-v_A t) + i\beta_n\right]\,,
\end{eqnarray}
where $k_1 = k_{min} = 2 \pi / \lambda_{\rm max}$ and $k_N = k_{max} = 2 \pi
/ \lambda_{\rm min}$ are, respectively, the wavenumbers corresponding to the maximum 
and minimum wavelengths associated with the turbulent field, the angle $\alpha_n$ 
and phase $\beta_n$ are randomly selected between $0$ and $2\pi$, and the random 
choice of $\pm$ selects the helicity of the wavevector about the $x'$ axis. The 
corresponding turbulent electric field is given by
\begin{eqnarray}
\delta {\bf E}({\bf r},t) = {mc\over q} {v_A\over c} 
\sum_{n = 1}^{N} \Omega_n 
\left[\pm i \sin \, \alpha_n
\hat y' - \cos\, \alpha_n \hat z' \right] \nonumber \\
\,\hbox{exp}\left[i k_n (x'-v_A t) + i\beta_n\right]\,.
\end{eqnarray}

The determination of the random polarization of each wavevector $k_n$ in the
laboratory frame is accomplished via the two-angle rotation matrix
\begin{eqnarray}
{\bf R} = \left(
 \begin{array}{ccc} 
\cos\theta_n & -\sin\theta_n \cos \phi_n & \sin \theta_n \sin \phi_n\\
 \sin \theta_n & \cos \theta_n \cos \phi_n & -\cos \theta_n \sin \phi_n\\
 0 & \sin \phi_n & \cos \phi_n
\end{array}
\right)
\label{eight}&\end{eqnarray}
where $ 0 \le \phi_n \le 2\pi$, and $0 \le \cos \theta_n \le 1$ are selected randomly
(for a total of five random components for each value of $n$).\footnote{The ZX rotation 
scheme adopted here differs from that presented in Giacalone \& Jokipii (1994).}
Throughout this work, the turbulent field structure at any position ${\bf r}$ is 
calculated by summing over $N = 25\,\log_{10}$[$\lambda_{\rm max} / \lambda_{\rm min}$]
values of wavevectors $k_n$, evenly spaced on a logarithmic scale between 
$k_{min}$ and $k_{max}$ (as justified in \S 5).  
Specifically, the 
particle position in the primed frame 
${\bf r'} = {\bf R}\cdot {\bf r}$ is used to calculate the 
real part of the turbulent 
magnetic field for each wavevector $k_n$, as given by
\begin{eqnarray}
Re\{\,\delta {\bf \Omega}({\bf r},t)_n'\} = 
\nonumber \\
\Omega_n \big\{\hbox{\rm cos} \,\alpha_n\,
\hbox{\rm cos}\,\left[ k_n \left( x'-v_A t\right)+\beta_n\right] \hat y' 
\nonumber \\ \pm  \hbox{\rm sin} 
\,\alpha_n \, \hbox{\rm sin} \, \left[ k_n \left( x'-v_A t\right)+\beta_n\right]
\hat z' \big\}\;.
\end{eqnarray}
Since each $k_n$ component is randomly oriented (i.e., has its unique value of 
$\hat y'$ and $\hat z'$), one must perform the rotation back to the unprimed frame  
$\delta {\bf \Omega(r)}_k = {\bf \tilde R}\cdot \delta {\bf \Omega(r)}_k'$
(where $\tilde {\bf R}\cdot {\bf R} = {\bf I}$---e.g., $\tilde R_{i,j} = R_{j,i}$) 
before performing the sum over $n$.

The desired spectrum of the turbulent magnetic field is set
through the appropriate choice of $\Gamma$ in the scaling
\begin{equation}
\Omega_n^2 = \Omega_1 ^2\left[{k_n \over k_1} \right]^{-\Gamma}
{\Delta k_n\over \Delta k_1} 
= \Omega_1^2\left[{k_n \over k_1} \right]^{-\Gamma+1}
\end{equation}
(as we have indicated, $\Gamma = 1$ for Bohm, 3/2 for Kraichnan, and 5/3 Kolmogorov),
where the quantity $\Omega_1$ is set by a parameter $\xi$ that specifies the energy 
density of the turbulent field via the definition 
\begin{equation}
\Omega_1^2 \sum_n \left[{k_n \over k_1}\right]^{-\Gamma+1}\,
 = \xi\, \Omega_0^2\;.
\end{equation} 
We note that for our adopted scheme, the value of $\Delta k_n / k_n$
is the same for all values of $n$.
We further note that $\xi = 2$ corresponds to the real part of the turbulent field
having the same energy density as a uniform field $\Omega_0$
since $\delta {\bf \Omega} \cdot
\delta {\bf \Omega}^* = 2\,Re\{\delta {\bf \Omega}\} ^2$.
Here we assume that there are a sufficiently large number of randomly
polarized transverse waves so that the cross terms of the above dot 
product cancel each other out.

\section{Dimensionless Equations of Motion}

The equations that govern the motion of relativistic charged particles
through the turbulent medium are
\be
{d {\bf u}\over  dt} = {q\over mc} \left( \delta {\bf E} +
{{\bf u} \times{\bf B}\over \gamma}\right)\;,
\ee
and
\be
{d {\bf r}\over dt} = {\bf v}\;,
\ee
where ${\bf u} = \gamma {\bf v}/c$ and $\gamma$ is the particle Lorentz factor.  
As can be seen from the form of Equations~(7) and (8), $\delta E$$\sim$$(v_A/c)\, 
\delta B$. Since MHD fluctuations in molecular clouds are expected to propagate at 
speeds of $v_A \sim 1$ km s$^{-1}$, $\delta E << \delta B$, and the electric field has a 
negligible effect on the local particle motion for particle speeds approaching $c$.  
However, electric fluctuations can significantly accelerate charged particles given a 
sufficiently long time (Fraschetti \& Melia 2008). Under the most ideal conditions, 
turbulent fields can energize protons in a time $\Delta t$ by an amount
\be
\Delta E_p = e \,\delta E  \,c\Delta t
\approx e \,\delta B\, v_A\Delta t \,.
\ee
Such an ideal acceleration, however, can only occur for time intervals 
$\Delta t < \lambda_{\rm max} / c$. For the parameter values adopted here 
($v_A = 1$ km s$^{-1}$, $\delta B = 10\;\mu$G, $\lambda_{\rm max} = 1$ pc), this 
ideal acceleration may only last for $\sim$$3$ yrs and energize particles 
by an amount $\Delta E_p \approx 0.03$ TeV. For longer time intervals, the
process becomes stochastic and the particle energy increases as $\Delta E_p 
\propto \sqrt{t}$. A reasonable upper limit to the increase in particle energy 
as a function of time is therefore given by
\be
\Delta E_{p;\; {\rm max}} \sim 0.01 \, {\rm TeV}\, 
\left({t \over 1 \, {\rm yr}}\right)^{1/2}\;.
\ee

In order to both confirm this result and to obtain a more exact value for $\Delta 
E_{p;\; {\rm max}}$, we have solved Equations~(13) and (14) for protons moving 
trough a turbulent field characterized by $\Gamma = 3/2$, $\lambda_{\rm max} = 1$ pc, 
$\lambda_{\rm min} = 10^{-4}$ pc, an energy density equal to that of a uniform $B_0 = 
10$ $\mu$G magnetic field, and our adopted fiducial value of $v_A$ = 1 km s$^{-1}$. 
Since the focus of our paper is on relativistic particles whose radius of gyration
\begin{eqnarray}
R_g = {\gamma m c^2 \over q B_0} =\qquad\qquad\quad \nonumber \\
1.08\times 10^{-4} \,{\rm pc} \,\left({E_p \over 1 \,{\rm TeV}}\right)
\left({A \over Z}\right) 
\left({B_0\over 10 \, \mu {\rm G}}\right)^{-1}
\end{eqnarray}
falls within the values of $\lambda_{\rm min}$ and $\lambda_{\rm max}$, we have solved 
the resulting equations of motion for both a $10^2$ TeV and a $10^3$ TeV proton.
The resulting change in energy $|\Delta E_p|$ as a function of time for both 
particles is shown in Figure~1, and clearly demonstrates a random-walk 
behavior (for which $|\Delta E_p| \propto \sqrt{t}$) with fluctuations 
superimposed.  In addition, we find that Equation~(16)---as represented
by the dashed line in Figure~1---provides a good upper limit for $|\Delta E_p|$.
Since we focus our discussion on particle energies in excess of 1 TeV and 
diffusion times less than $10^4$ years, this test calculation shows that we 
may justifiably ignore the effects of the electric field in our work.
\begin{figure}
\figurenum{1}
\begin{center}
{\centerline{\epsscale{0.90} \plotone{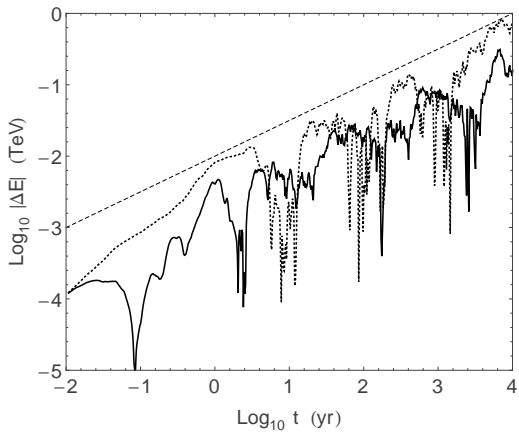} }}
\end{center}
\figcaption{The magnitude of the change in particle energy $|\Delta E_p|$
as a function of time for protons with initial energies of
$10^2$ TeV (solid curve) and $10^3$ TeV (dotted curve) moving 
through turbulent magnetic and electric fields characterized 
by $\Gamma = 3/2$, $\lambda_{\rm max} = 1$ pc, $\lambda_{\rm min}
 = 10^{-4}$ pc, and $v_A = 1$ km s$^{-1}$. The turbulent magnetic field
has an energy density equal to that of a uniform 10 $\mu$G field.
The dashed line represents the value of the upper limit 
$|\Delta E_{p;\; {\rm max}}|$ given by the expression in equation (16).}
\end{figure}

To simplify the analysis, we define a dimensionless time $\tau = t/t_0$, 
where $t_0$ is the inverse of the gyration frequency multiplied by
the Lorentz factor for a particle with 
charge $q = Ze$ and mass $m = A m_H$ in a reference field $B_0$, as given 
by the expression
\begin{eqnarray}
t_0 = {\gamma\over \Omega_0} = \qquad\qquad\qquad\qquad \nonumber \\ 
3.5\times 10^{-4}\, {\rm yrs}\,
\left({E_p\over 1\,{\rm TeV}}\right)
\left({A \over Z}\right) \left({B_0\over 10 \, \mu {\rm G}}\right)^{-1}\;.
\end{eqnarray}
We also define a corresponding dimensionless radius vector ${\bf \bar r}
= {\bf r} / R_g$. Since we ignore the electric field $\delta E$, $|{\bf u}| 
= \gamma v/c$ is a constant of the motion. Thus, for relativistic particles 
($v \approx c$), setting the value of $R_g$ also sets the value of $t_0$ 
(and vice versa) since $R_g = c t_0$.

Ignoring the electric field, the equations of motion for highly relativistic
particles
can then be written in dimensionless form as
\be
{d \hat {\bf u} \over  d\tau} = 
{\hat {\bf u} \times{\bf \bar B}}\;,
\ee
and
\be
{d {\bf \bar r}\over d\tau} = \hat {\bf u}\;,
\ee
where ${\bf \bar B} = {\bf B}/B_0$ 
and $\hat {\bf u} = {\bf u} / |{\bf u}|$.

\section{The Case of a Purely Turbulent Field}

In our formalism, the trajectory of a particle moving through a purely 
turbulent field is fully described by the four dimensionless parameters
$\Gamma$, $u_A = v_A/c$, $\bar\lambda_{\rm min} = \lambda_{\rm min}/R_g$ 
and 
$\bar\lambda_{\rm max} = \lambda_{\rm max}/R_g$
(related to the rigidity $\rho$ through the expression $\bar\lambda_{{\rm max}} = 2\pi/\rho$), 
along with the adopted
prescription for setting the $N$ values of wavevectors $k_n$
discussed below Equation~(9).  It is important to note that as the 
particle moves through the field, the radius of gyration changes 
depending on the field strength being sampled. Within this context, 
$B_0$ is taken to be the field strength of a uniform field whose 
energy density equals that of the turbulent field. In turn, the 
value of $R_g$ represents a characteristic value for a 
particle's radius of gyration.

We begin our analysis by considering how motion through a time-dependent 
turbulent field differs from that of a static turbulent field ($v_A = 0$).
To this end, we calculate the trajectory of a particle over a time $\tau_{max} = 
10^5 \,\bar\lambda_{\rm max}$ for the case $\Gamma = 3/2$ and the following four 
sets of wavelength ranges $[\bar\lambda_{\rm min}, \bar\lambda_{\rm max}]$: [3,300]; 
[0.3,30]; [0.03,3]; and [0.003,0.3]. We plot the displacement $\bar r$ of each 
particle as a function of (the dimensionless) time $\tau$ in Figure~2 for the 
case of a static field ($v_A = 0$), and in Figure~3 for the case of a 
time-dependent magnetic field with an adopted fiducial value $v_A$ = 1 km s$^{-1}$. 
The long-dashed lines serve as a reference and have slopes of 1/2.  
\begin{figure}
\figurenum{2}
{\centerline{\epsscale{0.90} \plotone{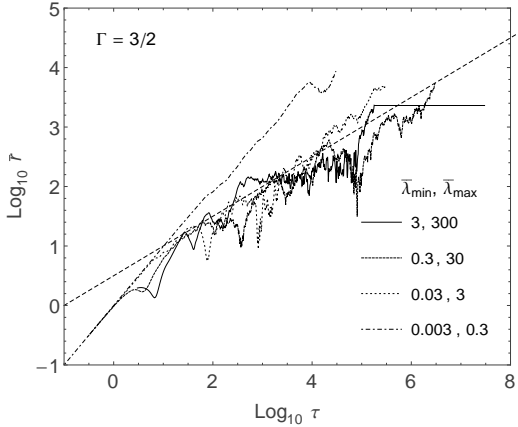} }}
\figcaption{The displacement $\bar r$ as a function of time $\tau$ for four 
particles moving through a static ($v_A = 0$) turbulent magnetic field with 
index $\Gamma = 3/2$. The values of $\bar\lambda_{\rm min}$ and $\bar\lambda_{\rm max}$
correspond to the following curves: 3, 300 (solid); 0.3, 30 (short-dashed); 
0.03, 3 (dotted); 0.003, 0.3 (dot-dashed). The long-dashed line 
serves as a reference and has a slope of 1/2.} 
\end{figure}
\begin{figure}
\figurenum{3}
{\centerline{\epsscale{0.90} \plotone{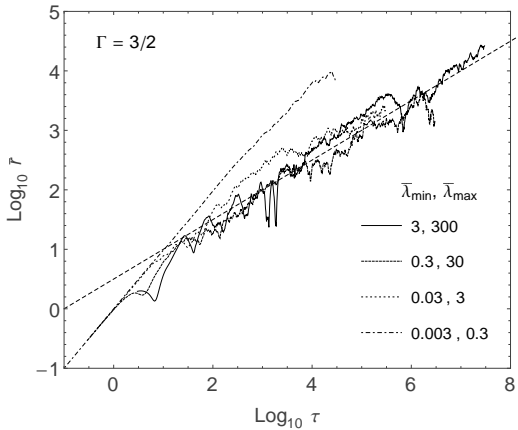} }}
\figcaption{The displacement $\bar r$ as a function of time $\tau$ for four 
particles moving through a temporally fluctuating ($v_A = 1$ km s$^{-1}$) turbulent magnetic 
field with index $\Gamma = 3/2$. The values of $\bar\lambda_{\rm min}$ and $\bar\lambda_{\rm max}$
correspond to the following curves: 3, 300 (solid); 0.3, 30 (short-dashed); 
0.03, 3 (dotted); 0.003, 0.3 (dot-dashed). The long-dashed line 
serves as a reference and has a slope of 1/2.} 
\end{figure}

Figures 2 and 3 illustrate three important points. First, particles with a radius of 
gyration below the range of turbulent wavelengths may eventually get trapped 
in a static field, as can be seen by the fact that $\bar r$ is constant 
at times $\tau > 10^5$ for the $\bar\lambda_{min} = 3$ particle (solid line in Figure 2).
To gain insight into this phenomenon, we plot in Figure 4 the dot product $\hat {\bf B} \cdot \hat {\bf v}$ 
as a function of time for the trapped particle shown in Figure 2 (solid curve).  One sees that
trapping occurs when particles move nearly perpendicular to the local magnetic field, oscillating in a
sort of local magnetic bottle.  
As can be seen in Figure~3 from the solid line at  times  $\tau > 10^5$, time-dependent 
fluctuations will disrupt this trapping on an expected timescale $\tau_{MHD} \sim  \bar\lambda_{\rm min}/u_A$
($\sim 10^6$ for the solid curve shown in Figure~3). Second, 
once particles with radii of gyration smaller than $\lambda_{\rm max}$ have moved beyond a 
(dimensionless) distance $\sim \bar\lambda_{\rm max}$, their displacement scales as $\bar r 
\propto \tau^{1/2}$. Finally, particles with a radius of gyration greater than the maximum 
turbulent wavelength are not strongly affected by local turbulence. The motion of such 
(highly-energetic) particles will not be considered in our analysis.  
\begin{figure}
\figurenum{4}
{\centerline{\epsscale{0.90} \plotone{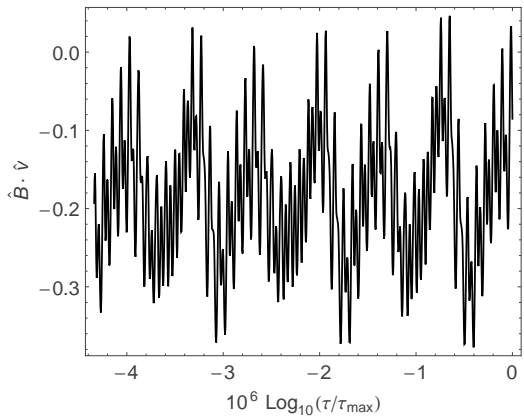} }}
\figcaption{The dot product between the field
direction and particle direction of motion 
as a function of $\tau$ for the trapped particle
shown in Figure~2 (solid curve).} 
\end{figure}

The motion of charged particles through a turbulent
magnetic field is chaotic in nature.  As such, 
a complete analysis requires a statistical approach.
We have therefore performed a suite of experiments designed
to adequately sample our parameter space.  Specifically,
each experiment is defined by a choice of the parameters
$\Gamma$, $\bar\lambda_{\rm min}$, and $\bar\lambda_{\rm max}$.
We adopt the value of $v_A$ = 1 km s$^{-1}$, although 
in the absence of particle trapping, our results will not
be sensitive to this chosen value. 
For each run, we calculate the trajectory of
$N_p$ particles injected randomly from the origin
for a time $\tau_{max}$, with
each particle sampling its own unique magnetic field structure
(i.e., the values of $\alpha_n$, $\beta_n$, $\theta_n$, 
$\phi_n$ and the choice of a $\pm$ are 
chosen randomly for each particle).  
The suite of experiments performed for the case of a purely
turbulent field are summarized in Table 1. 

We plot the distributions of $\bar x = x/R_g$ and 
$\bar r = r/R_g$ at time $\tau = 10^3 \,\bar\lambda_{\rm max}$
for experiment 2 in Figures~5--6.  (The corresponding
distributions for experiments 1 and 3 are qualitatively
very similar.)
Since the particles at this time have fully sampled
the turbulent structure of the field, the distributions
of their positions $\bar x$, $\bar y$ and $\bar z$ are
expected to be normal.  For a purely turbulent field, 
all three distributions are expected to have mean values
of zero and equal variances (within the expected statistical 
fluctuations).  
Furthermore, since motion along
any axis is independent of the others, then the 
displacement vector $\bar r = \sqrt{\bar x^2 +\bar y^2
+\bar z^2}$ has three independent orthogonal components,
each of which follow a standard normal distribution.
As such, the $\bar r$ values should be distributed
according to a chi distribution with 3 degrees of freedom.
To illustrate these points,  we include the corresponding Gaussian curve derived from 
the mean and variance in Figure~5, and the corresponding $k = 3$ chi distribution
in Figure~6.  As illustrated by 
our results, cosmic-ray diffusion through turbulent
magnetic fields is well represented
by Gaussian statistics.  
\begin{figure}
\figurenum{5}
{\centerline{\epsscale{0.90} \plotone{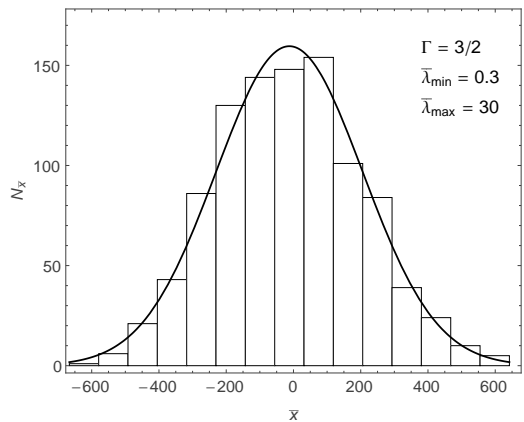} }}
\figcaption{The distribution of $\bar x$ values
at time $\tau = 10^3 \,\bar\lambda_{\rm max}$ for
experiment 2 (histogram), superimposed with a 
Gaussian function (black curve) with the same
mean and variance.} 
\end{figure}
\begin{figure}
\figurenum{6}
{\centerline{\epsscale{0.90} \plotone{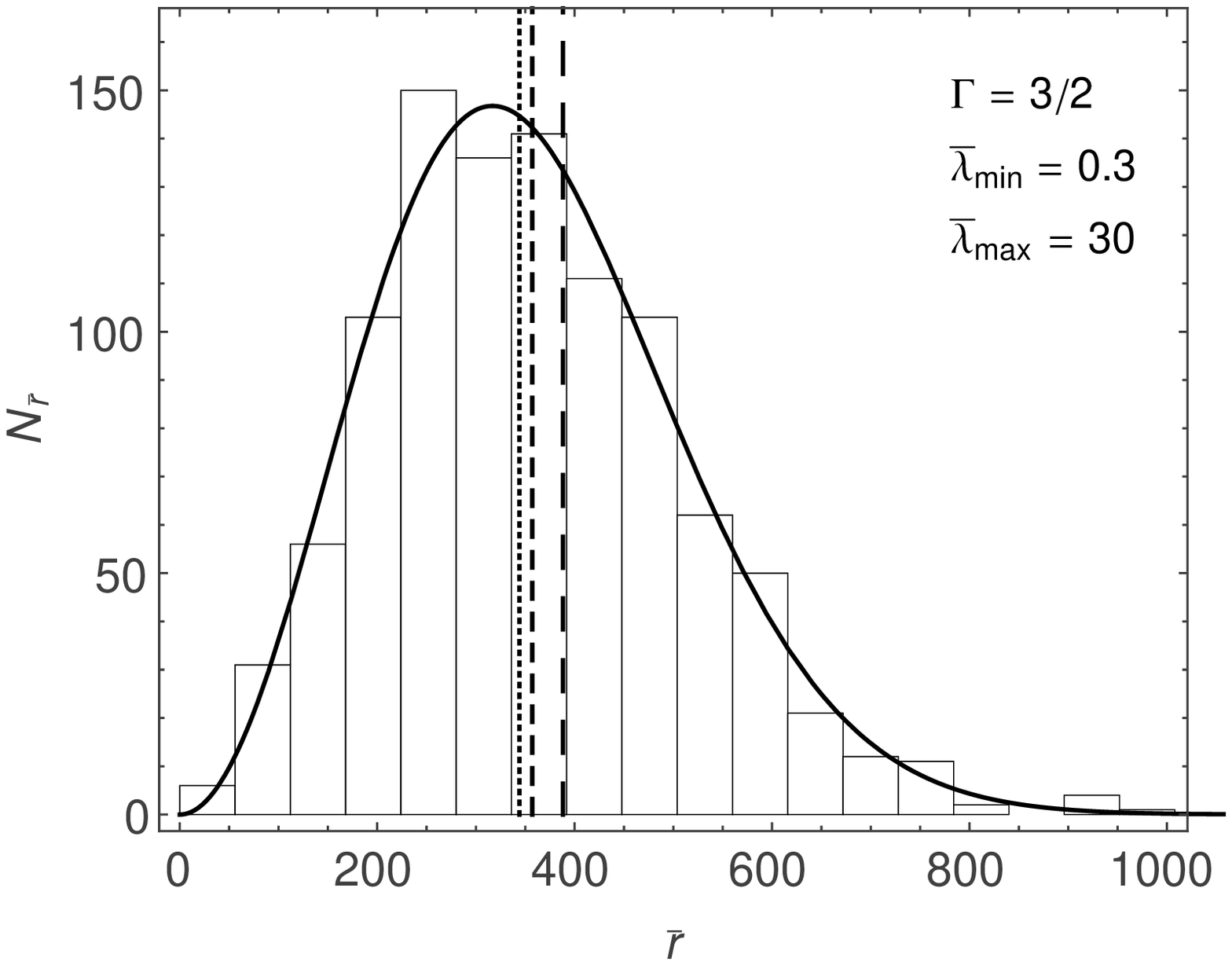} }}
\figcaption{The distribution of $\bar r$ values
at time $\tau = 10^3\, \bar\lambda_{\rm max}$ for
experiment 2 (histogram), superimposed with a 
chi function of degree 3 (black curve) and scaled
using the mean of the $x$, $y$ and $z$ distribution
variances.  The vertical dotted, short-dashed, and long-dashed
lines represent the median, mean and rms values 
for the distributions, respectively.} 
\end{figure}

The median, mean and rms values of the $\bar r$
distribution shown in Figure 6 are denoted, respectively, 
by the vertical dotted, short-dashed, and long-dashed 
lines.  Although each of these output measures characterize 
the distribution, we will adopt the mean value $\langle \bar r \rangle$ of the 
particle  displacements  as our primary 
output measure, and calculate its  value at several times $\tau$
for each experiment performed (as listed in Table 1).
In order to determine how sensitive
the value of our output measure is on $\bar\lambda_{\rm min}$,
we compare the results of experiments 1--3 with those of
experiments 7, 9 and 11
in Figure 7.  As clearly illustrated by the overlap between
the results from experiments 1 (open triangle) and 7 (solid triangle), 2 
(open square) and 9 (solid square), and 3 (open circle) and 11 (closed circle), 
particle diffusion depends primarily on the maximum turbulence
wavelength $\bar\lambda_{\rm max}$, and is not 
sensitive to the minimum turbulence wavelength $\bar\lambda_{\rm min}$,
so long as the radius of gyration is greater than the minimum
turbulence wavelength (see discussion in \S 7).
Our analysis is therefore greatly simplified in that 
there is only one primary parameter -- $\bar\lambda_{\rm max}$ -- that dictates
how particles diffuse through a purely turbulent 
field with a specified value of $\Gamma$. We also 
note that the values of $\langle \bar r \rangle$
clearly exhibit the $\tau^{1/2}$ dependence associated with
a diffusion process (though particles
with $R_g \sim \lambda_{\rm max}$ have motions 
intermediary to their counterparts
with smaller radii of gyration and the free-streaming
motion of their counterparts with greater radii of gyration).
\begin{figure}
\figurenum{7}
{\centerline{\epsscale{0.90} \plotone{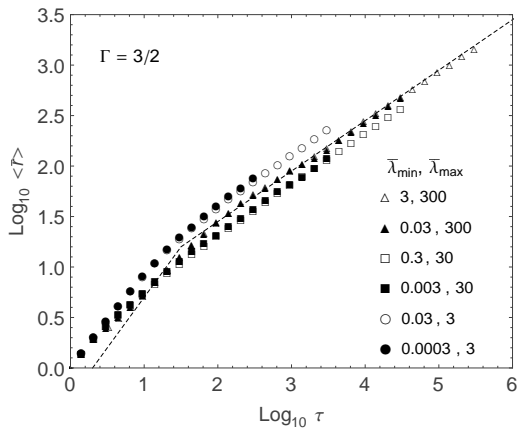} }}
\figcaption{The value of $\langle \bar r \rangle$ as a function of
$\tau$ for experiments 1 (open triangle), 2 (open square), 3 
(open circle), 7 (solid triangle), 9 (solid square) and 11 (solid circle).
The dashed lines serve as a reference and have slopes of 1/2 and 1.} 
\end{figure}

We focus the rest of our analysis on cases for which the 
particle gyration radius falls comfortably within the range
of the maximum and minimum turbulence wavelengths so that
particles undergo actual diffusion---that is, for which 
$\bar\lambda_{\rm max} \gg 1 \gg \bar\lambda_{\rm min}$.
To do so, we consider a turbulent field with a dynamic
range in wavelengths that span either four or five orders of magnitude.
We note, however, that the minimal dependence that particle diffusion has on
the smallest wavelength implies that our results can be 
extrapolated to lower values of $\bar\lambda_{\rm min}$
(see discussion in \S 7).  

A fundamental issue in this analysis is what value of $N$
will allow our discrete treatment of the turbulent field to
adequately represent a continuous field.  Toward that end,
we first note that the variance of the mean values of $\langle
\bar r \rangle$ is given by $\sigma_{mean} = \sigma_{\bar r}/\sqrt{N_p}$.
Based on the results presented in Figure 6, 
$\sigma_{\bar r} \sim \langle \bar r \rangle / 2$, so that the calculated
mean of our sample population with $N_p = 200$
is expected to be within $3 \sigma_{mean} = 1.5 \langle\bar r\rangle / \sqrt{N_p} 
\approx 0.1 \langle\bar r\rangle$ 
of the true (parent) value with $\sim 99$\% confidence.  
We next perform experiments 6, 16 and 25 with values of $N$ = 50, 100, 200
and 300.  The resulting values of $\langle\bar r\rangle$ at time $\tau_{max}$ as a 
function of $N$ are
shown in Figure 8, where the error bars represent
the expected $3\sigma$ statistical error of $0.1 \langle\bar r\rangle$.  These results appear
to justify our adoption of  $N = 25\,\log_{10}$[$\lambda_{\rm max} / \lambda_{\rm min}$]
presented in \S2.
\begin{figure}
\figurenum{8}
{\centerline{\epsscale{0.90} \plotone{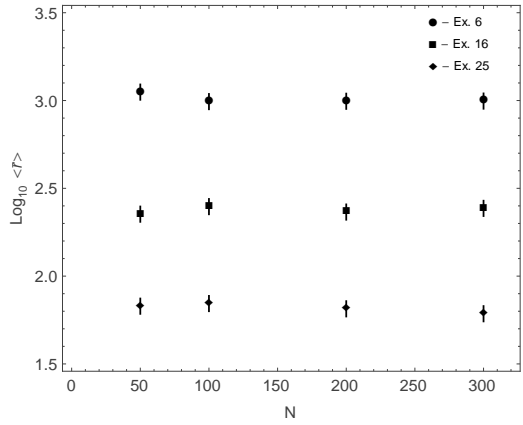} }}
\figcaption{The value of $\langle \bar r \rangle$ at $\tau_{max}$
as a function of
$N$ for experiments 6, 16, and 25.
The error bars represent
the expected $3\sigma$ statistical error of 10\%.} 
\end{figure}

The results of experiments 4--10, 12--18 and 19--25 
are presented in Figures~9, 10 and 11, respectively.  
A self-similar pattern is clearly visible in these 
figures for cases with $\bar\lambda_{\rm max} \ga 30$, with 
a break in the slope of the curves from $\sim 1$ to $1/2$ 
occurring around $\tau \sim\bar\lambda_{\rm max} /10$
for $\Gamma = 3/2$ and $5/3$, and at $\tau \sim 10$ for
$\Gamma = 1$.  We note, however, that the break is not
smooth for  the solid circles show in Figures 9, 11 and 12.
These irregularities  occur as particles with small radii or 
gyration make transitions from weakly perturbed propagation 
(for which $\langle\bar r \rangle\propto \tau$) to diffusion (for which
$\langle\bar r\rangle \propto \tau^{1/2}$).  
This feature indicates
 that as particles with small radii of gyration make this transition
 after traveling a distance $\sim 0.1 \lambda_{max}$, 
 they are effectively ``scattered'' randomly in all directions, so that on 
 average, their distance from the origin does not change appreciably 
 until they truly reach the diffusion regime (i.e.  they have been ``scattered'' 
 numerous times).    

\begin{figure}
\figurenum{9}
{\centerline{\epsscale{0.90} \plotone{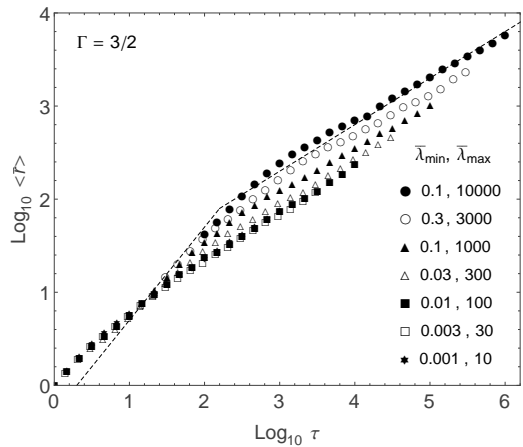} }}
\figcaption{The value of $\langle \bar r \rangle$ as a function of
$\tau$ for experiments 4--10, for which $\Gamma = 3/2$.
The dashed lines serve as a reference and have slopes of 1/2 and 1.} 
\end{figure}
\begin{figure}
\figurenum{10}
{\centerline{\epsscale{0.90} \plotone{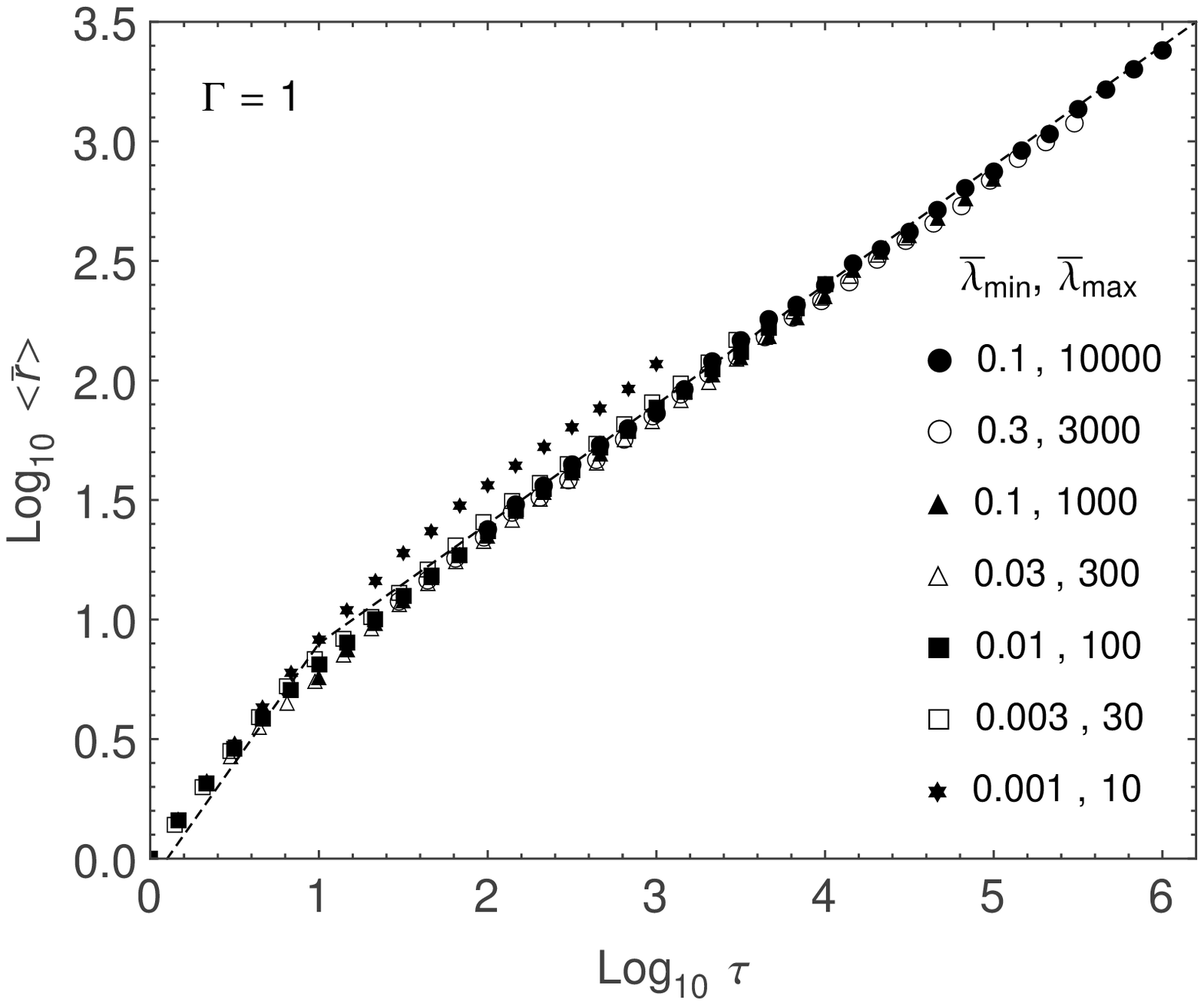} }}
\figcaption{The value of $\langle \bar r \rangle$ as a function of
$\tau$ for experiments 12--18, for which $\Gamma = 1$.
The dashed lines serve as a reference and have slopes of 1/2 and 1.} 
\end{figure}
\begin{figure}
\figurenum{11}
{\centerline{\epsscale{0.90} \plotone{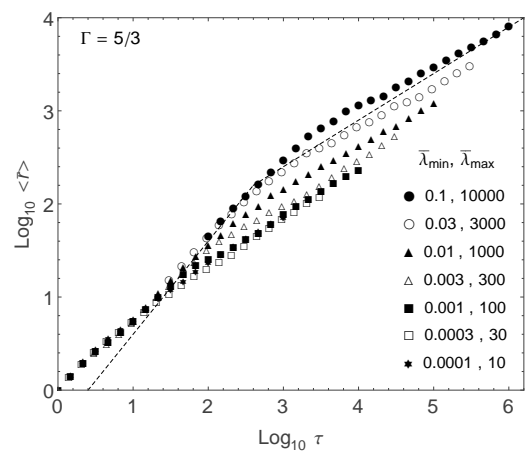} }}
\figcaption{The value of $\langle \bar r \rangle$ as a function of
$\tau$ for experiments 19--25, for which $\Gamma = 5/3$.
The dashed lines serve as a reference and have slopes of 1/2 and 1.} 
\end{figure}

In order to put our results into a physical context, 
we consider relativistic protons moving through a purely
turbulent magnetic field for
which $\lambda_{\rm max} = 1$ pc, and dimensionalize the results 
of experiments 4--10 accordingly through a proper choice
of $R_g = \lambda_{\rm max}/\bar\lambda_{\rm max}$.  
We note that setting a common value of $\lambda_{\rm max}$
for experiments 4--25 also sets a common value
of $t_{max} = \tau_{max} \, t_0 = 10^2 
\lambda_{\rm max} / c$. The results are 
presented in Figure~12. As previously noted, the solutions 
are nearly self-similar for particles whose radius of 
gyration is $R_g \la 0.03 \lambda_{\rm max}$. 
\begin{figure}
\figurenum{12}
{\centerline{\epsscale{0.90} \plotone{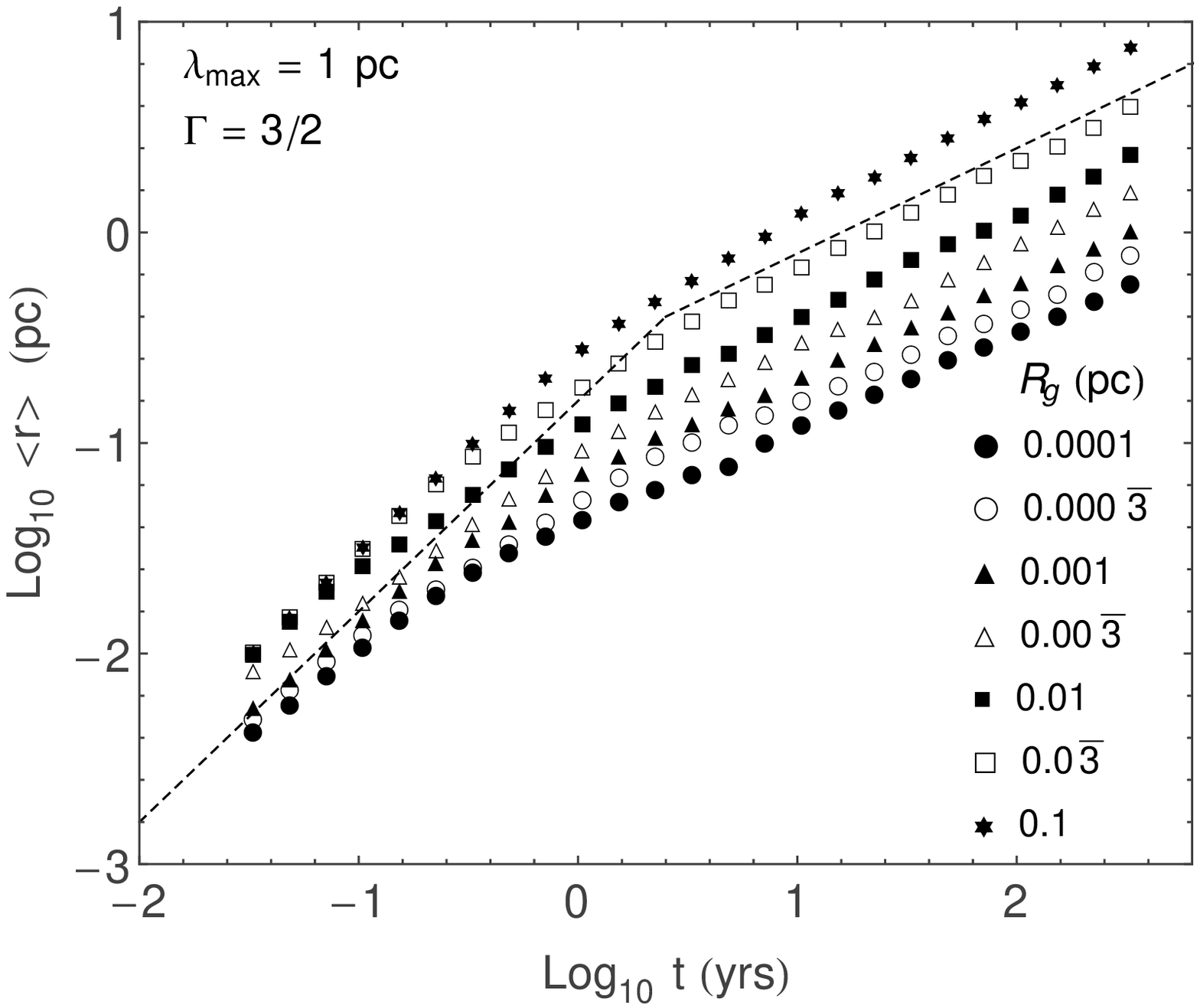} }}
\figcaption{The value of $\langle r \rangle$ as a function of
time for experiments 4--10, for which $\Gamma = 3/2$.  
The results of these experiments are dimensionalized 
by assuming that $\lambda_{\rm max} = 1$ pc for each case,
and setting the value of $R_g$ accordingly.  
The dashed lines serve as a reference and have slopes of 1/2 and 1.} 
\end{figure}

To better understand how a particle's gyration radius
helps determine the nature of its motion, we plot in 
Figure~13 particle trajectories of three
particles with different radii of gyration, each
injected with identical velocity from the origin into the
same turbulent (but static) magnetic field defined by $\Gamma = 3/2$,
$\lambda_{\rm max} = 1$ pc, $\lambda_{\rm min} = 
10^{-4}$ pc, and $B_0 = 10$ $\mu$G.  The field line that passes through
the origin is depicted by the thin black line.  Particle
trajectories are depicted by 
the blue ($R_g =  0.001$ pc),
green ($R_g =  0.01$ pc) and red ($R_g =  0.1$ pc)
curves.  Clearly, the nature of particle motion 
differs for particles with $R_g \la 0.01 \lambda_{\rm max}$
and $R_g \ga 0.01\lambda_{\rm max}$.  For the former,
particles are strongly coupled to field lines and 
their motion is directly tied to the field line structure, 
whereas for the latter,  particles
``random walk" through the field.
That is not to say that particles with small radii of gyration
move smoothly along field lines.
Rather, although they are scattered by the turbulent magnetic
fields according to their energies, their spread due to scatter is 
small compared to how far they propagate in the direction of the field.
\begin{figure}
\figurenum{13}
{\centerline{\epsscale{0.90} \plotone{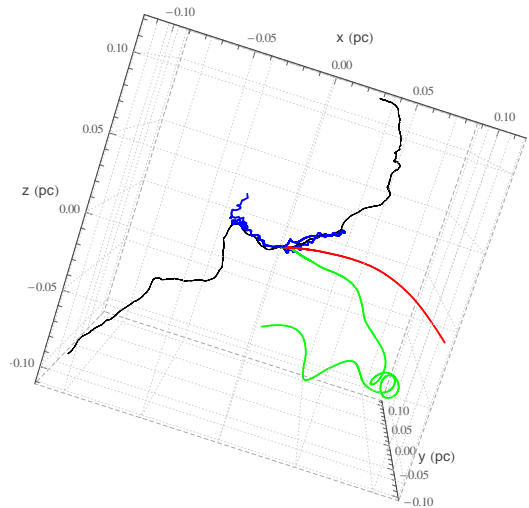} }}
\figcaption{Trajectories of three particles injected 
with identical velocities from the origin into the same turbulent
(but static) magnetic field, defined by $\Gamma = 3/2$,
$\lambda_{\rm max} = 1$ pc, and $\lambda_{\rm min} = 
10^{-4}$ pc.  The colored curves denote the path
of particles with gyration radii $0.001$ pc
(blue), $0.01$ pc (green),  and $0.1$ pc (red).  The black curve
denotes the magnetic field line passing through the origin.} 
\end{figure}

A central aspect of this work is a determination of
the relation between particle diffusion and energy.
To that end, we define a dimensionless energy
$\epsilon = E_p / E_0$, where
\begin{eqnarray}
E_0 = \lambda_{\rm max}\,e\,B_0 = \qquad\qquad\qquad\nonumber\\
9.2\times 10^3 \, {\rm TeV}
\,\left({\lambda_{\rm max} \over 1 \,{\rm pc}}\right)
\left({B_0\over 10\,\mu{\rm G}}\right)\,,
\end{eqnarray}
which then yields the relation 
$\epsilon = Z\,R_g / \lambda_{\rm max}= Z\,\bar\lambda_{\rm max}^{-1}$.
We plot the values of $\langle r \rangle/\lambda_{\rm max}$ 
at $\tau_{max}$ as a function of $\epsilon$ in Figure~14
for experiments 4--10, 12--18, and 19--25. Each set of results 
for a given value of $\Gamma$ demonstrates a clear break 
at $\eps_b\sim 0.005$, corresponding to particles with gyration 
radii $R_g \sim 0.005\lambda_{\rm max}/Z$. There is clearly a stronger dependence 
between $\langle r \rangle$ and $\eps$ above the break, presumably due to the 
fact that particles with $\eps << \eps_b$ are strongly coupled to 
the field lines, as shown in Figure~13.  As such, their diffusion 
is dictated primarily by the field structure, and hence, becomes 
less sensitive to their energy/radius of gyration. 
Specifically, particles with radii of gyration smaller than $\sim  0.005 \lambda_{max}$ 
will effectively scatter off field fluctuations that have a similar length scale as their gyration radius.  
In contrast, particles with sufficiently large gyration radii effectively decouple from the field-lines 
(as is illustrated in Figure 13),  and essentially random walk through the field on length 
scales equal to their gyration radius.  Their motion, therefore, is not very sensitive 
to the nature of the small-scale fluctuations, as can be seen by the convergence 
of the output values in this regime for the $\Gamma = 5/3$ and $\Gamma = 1/2$ cases.
\begin{figure}
\figurenum{14}
{\centerline{\epsscale{0.90} \plotone{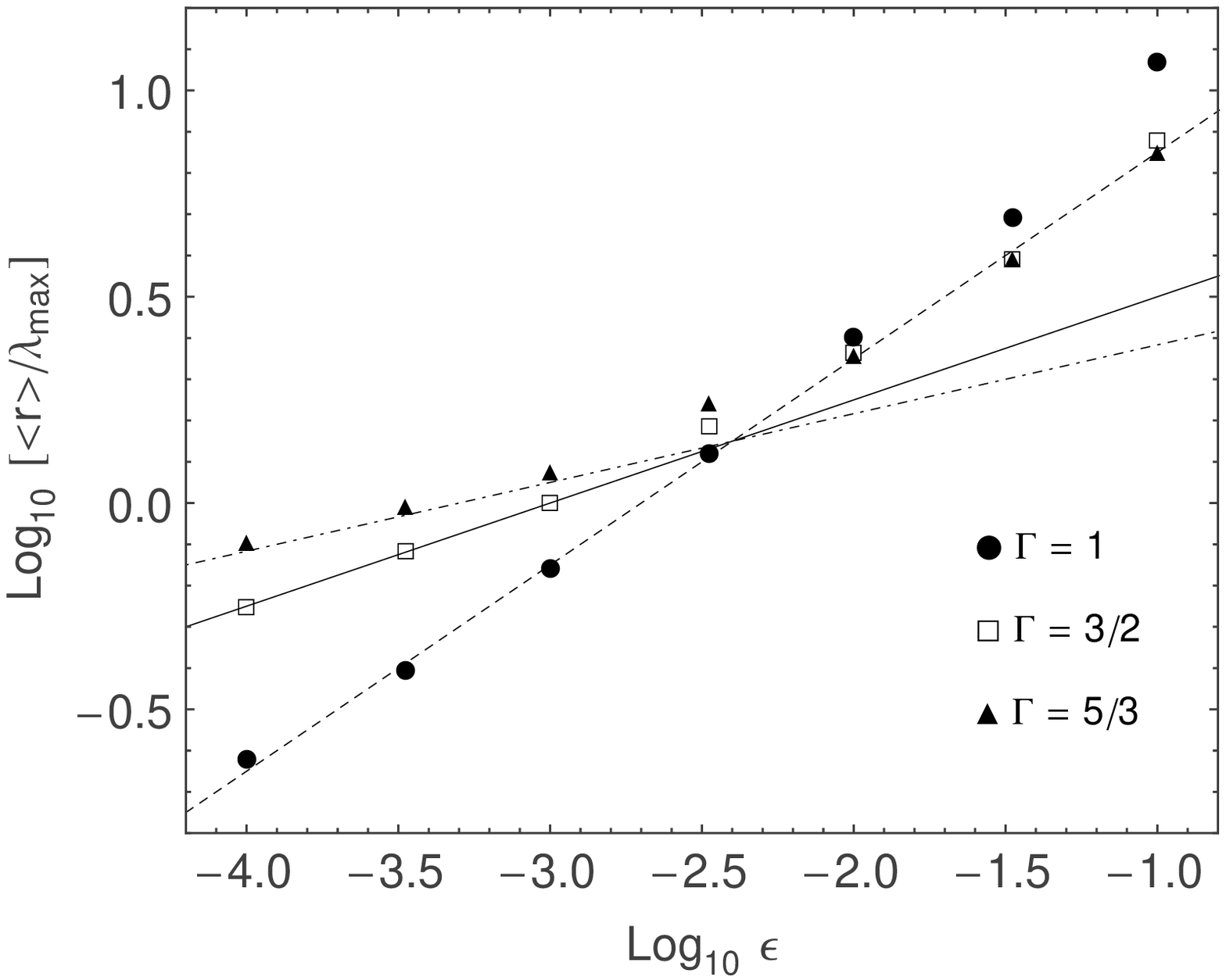} }}
\figcaption{The value of $\langle r \rangle / \lambda_{\rm max}$ evaluated
at $\tau_{max}$ as a function of
$\eps$ for experiments 4--10 ($\Gamma = 3/2$), 12--18
($\Gamma = 1$), and 19--25 ($\Gamma = 5/3$).
The dot-dashed ($\Gamma = 5/3$), solid 
($\Gamma = 3/2$) and dashed ($\Gamma = 1$) curves represent power-law fits to the
data, as discussed in the text.} 
\end{figure}

In order to put our results into a useful format, we note that in the standard
theory for particle diffusion, the turbulent field index $\Gamma$ is related to the 
diffusion coefficient index $\delta$ (as defined in Equation 2) through the
expression $\delta = 2-\Gamma$.  As such, the diffusion radius 
$R_{\rm diff} \propto E_p^{1-\Gamma/2}\,t^{1/2}$.  In turn, we
express the particle diffusion length as a function of energy and time
through the expression
\be
{\langle r \rangle} = \lambda_{\rm max}\,\Lambda \left({E_p\over E_0}\right)^{\alpha}
\,\left({t \over t_c}\right)^{1/2}\,,
\ee
where
\be
t_c = {\lambda_{\rm max}\over c} = 3.3 \,{\rm yrs}
\left({\lambda_{\rm max}\over 1\,{\rm pc}}\right)\,.
\ee
We then fit the three lowest-energy data points for each case
shown in Figure 14 at time $t = 100\, t_c$, 
as illustrated by the dashed ($\Gamma = 1$), solid ($\Gamma = 3/2$)
and dash-dotted ($\Gamma = 5/3$) lines, where 
the corresponding values of $\Lambda$ and $\alpha$ 
are given in Table 3 for each value of $\Gamma$.  In all cases,
good fits are obtained with $\alpha = 1-\Gamma/2$ for 
$\eps \la 0.005$.

\section{Uniform Field with a Turbulent Component}
We next consider a molecular cloud environment
threaded by a uniform magnetic field with a strong
turbulent component.  Specifically, we assume
a magnetic field of the form ${\bf B} ({\bf r},t) 
= B_0 \hat z +\delta {\bf B} ({\bf r},t)$. 
In our formalism, the motion of a particle moving through 
such a field is then described by five dimensionless
parameters:  $\Gamma$, $u_A$, $\bar\lambda_{\rm min}$,
$\bar\lambda_{\rm max}$ and $\eta$.  

Observations of molecular clouds suggest that the
magnetic fluctuations have amplitudes $\delta B
\sim B_0$.  This finding follows from considering
the observed non-thermal line-widths in molecular clouds
(Larson 1981; Myers et al 1991) to result from 
MHD waves (e.g., Fatuzzo \& Adams 1993; McKee \& 
Zweibel 1995; see Fatuzzo \& Adams 2002 for further
discussion). 
We therefore consider the case that the magnetic 
energy density of the turbulent field equals that
of the underlying field, thereby setting $\xi = 2$
for all cases explored. The suite of experiments
performed are summarized in Table 2.

The introduction of the field $B_0 \hat{z}$ has broken the
isotropy, so we now plot both the distribution of $\bar x = x/R_g$ and 
that of $\bar z = z/R_g$ at time $\tau = 10^2 \,\bar\lambda_{\rm max}$
for experiment 5 (Table 2) in Figures~15 and 16, were the solid curves 
depict the corresponding Gaussians derived from the mean and variance 
of each distribution. As illustrated by 
our results, cosmic-ray diffusion through uniform magnetic fields
with strong turbulent components is fairly well represented
by Gaussian statistics.
\begin{figure}
\figurenum{15}
{\centerline{\epsscale{0.90} \plotone{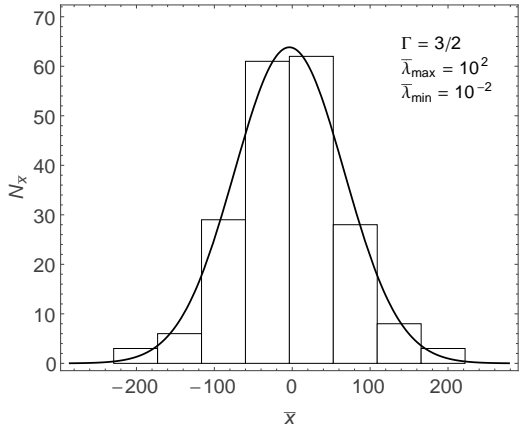} }}
\figcaption{The distribution of $\bar x$ values
at time $\tau = 10^2 \bar\lambda_{\rm max}$ for
experiment 5 in Table 2 (histogram), superimposed with a 
Gaussian function (black curve) with the same
mean and variance.} 
\end{figure}
\begin{figure}
\figurenum{16}
{\centerline{\epsscale{0.90} \plotone{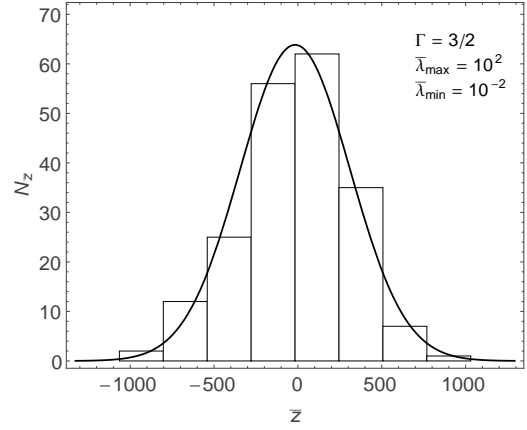} }}
\figcaption{The distribution of $\bar z$ values
at time $\tau = 10^2 \bar\lambda_{\rm max}$ for
experiment 5 in Table 2 (histogram), superimposed with a 
Gaussian function (black curve) with the same
mean and variance.} 
\end{figure}

The rms values of the particle positions $\bar x$ and $\bar z$ at several 
times $\tau$ for each experiment are shown in Figures~17 and 18. 
As found for the purely turbulent field discussed in 
\S 5,  the curves appear to be nearly self-similar, 
with a break in the slope of the curves
occurring at around $\tau \sim\bar\lambda_{\rm max}/ 10$.  
Not surprisingly, particles diffuse further along
the direction of the uniform field than they do across the field,
with $\bar z_{rms} \sim 5 \bar x_{rms}$.
\begin{figure}
\figurenum{17}
{\centerline{\epsscale{0.90} \plotone{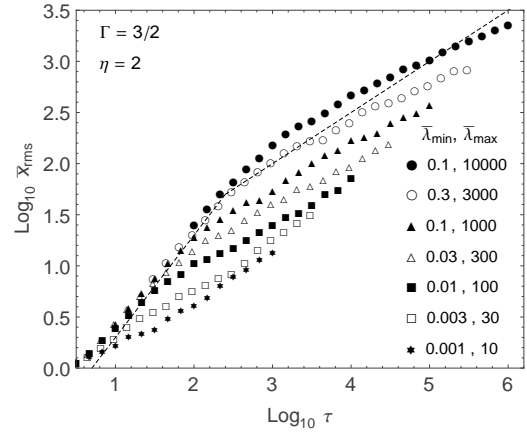} }}
\figcaption{The value of $\bar x_{rms}$ as a function of
$\tau$ for experiments 1--7 listed in Table 2.  The dashed lines
serve as a reference and have slopes of 1/2 and 1.} 
\end{figure}
\begin{figure}
\figurenum{18}
{\centerline{\epsscale{0.90} \plotone{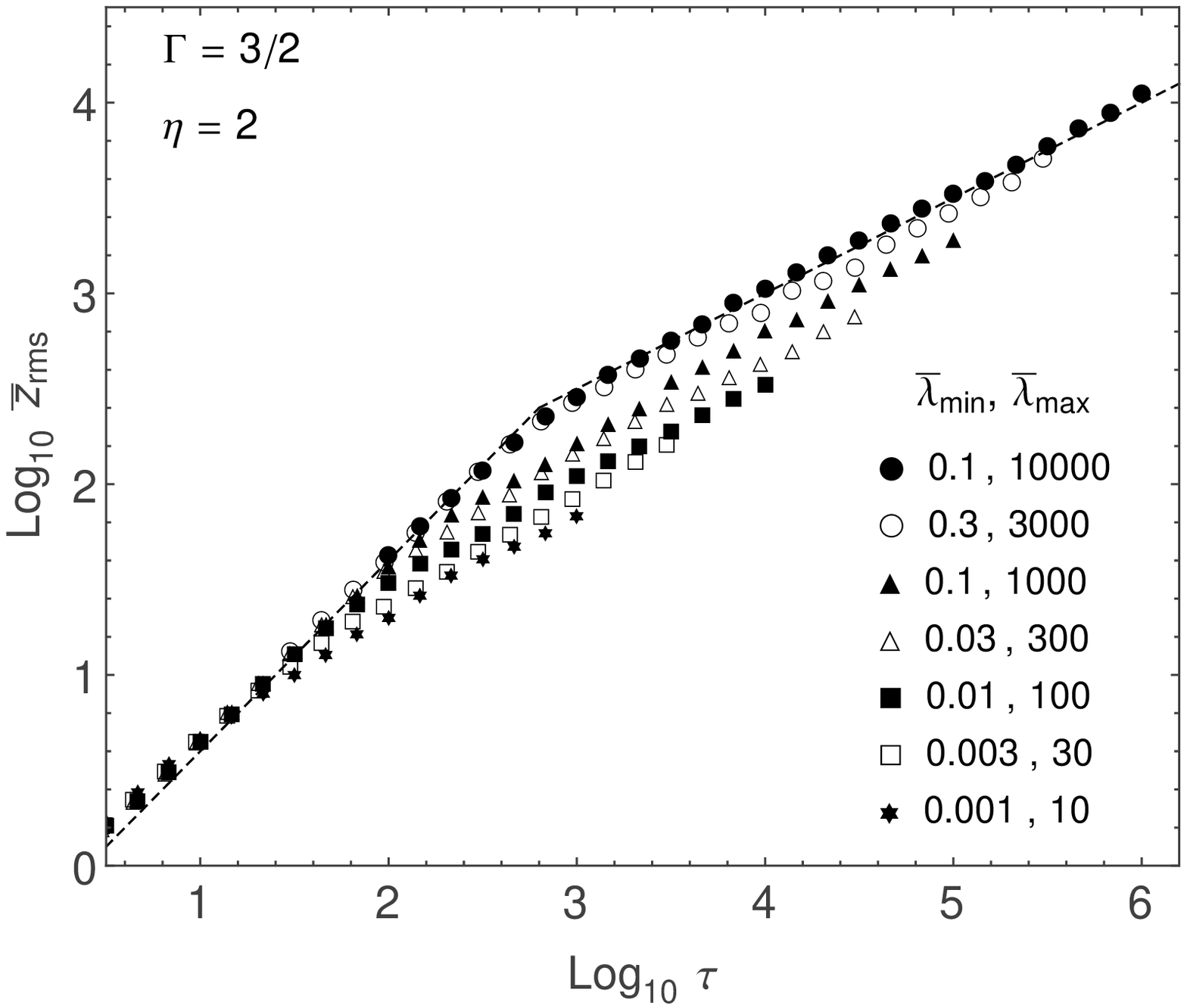} }}
\figcaption{The value of $\bar z_{rms}$ as a function of
$\tau$ for experiments 1--7 listed in Table 2. The dashed lines
serve as a reference and have slopes of 1/2 and 1.} 
\end{figure}

As noted in \S 5, a central aspect of this work is a determination of
the relation between particle diffusion and energy.
To that end, we plot the values of $x_{rms}/\lambda_{\rm max}$ 
and $z_{rms}/\lambda_{\rm max}$ at $\tau_{max}$ as a function
of $\epsilon$ in Figure~19 for experiments 1--21 listed in Table 2.
In all cases, the data for diffusion along the underlying magnetic  
field direction is well-fit by a line.
Likewise, the data for the diffusion across the underlying magnetic field
is well-fit by a line for $\Gamma = 1$ and $\Gamma = 3/2$, but does exhibit
a break at $\eps\sim 0.01$ for $\Gamma = 5/3$.

Following the analysis presented in \S 5, we
express the particle diffusion lengths across and along the 
underlying uniform magnetic field 
through the expressions
\be
x_{rms} = \lambda_{\rm max}\,\Lambda_x \left({E_p\over E_0}\right)^{\alpha_x} 
\,\left({t \over t_c}\right)^{1/2}\,,
\ee
and
\be
z_{rms} = \lambda_{\rm max}\,\Lambda_z \left({E_p\over E_0}\right)^{\alpha_z} 
\,\left({t \over t_c}\right)^{1/2}\,.
\ee
We fit the data in Figure 19 at time $t = 100\, t_c$, 
as illustrated by the dashed ($\Gamma = 1$), solid ($\Gamma = 3/2$)
and dash-dotted ($\Gamma = 5/3$) lines, where 
the corresponding values of $\Lambda$ and $\alpha$ 
are given in Table 4 for each value of $\Gamma$.  In all cases except
for $x_{rms}$ when $\Gamma = 1$,
good fits are obtained with $\alpha = 1-\Gamma/2$ for the entire range of
$\eps$ explored.

\begin{figure}
\figurenum{19}
{\centerline{\epsscale{0.90} \plotone{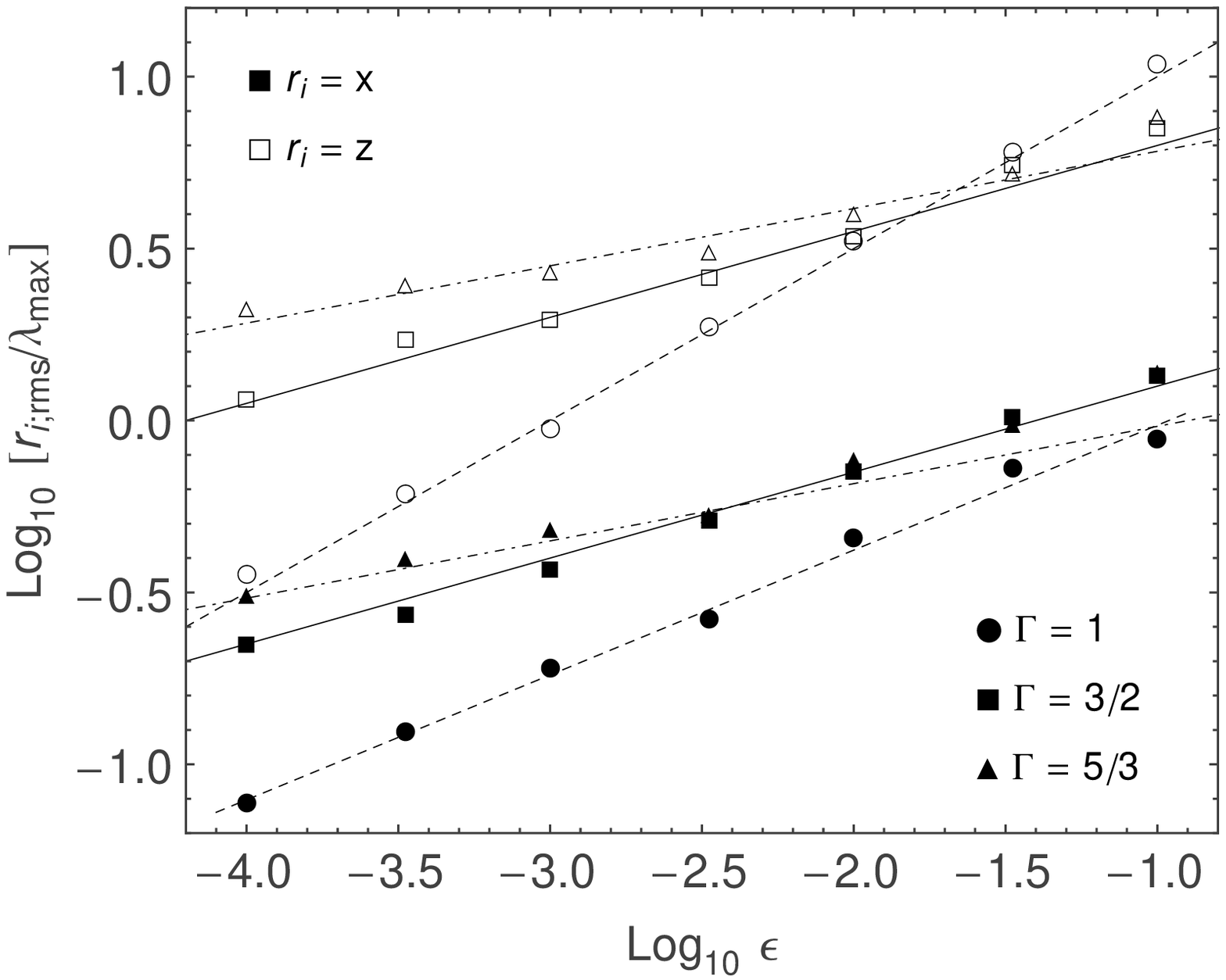} }}
\figcaption{The values of $x_{rms} /\lambda_{\rm max}$ (solid)
and $z_{rms} /\lambda_{\rm max}$ (open) evaluated
at $\tau_{max}$ as a function of
$\eps$ for experiments 1--7 ($\Gamma = 3/2$), 8--14
($\Gamma = 1$), and 15--21 ($\Gamma = 5/3$) listed in Table 2.
The dot-dashed ($\Gamma = 5/3$), solid 
($\Gamma = 3/2$) and dashed ($\Gamma = 1$) curves represent fits to the
data, as discussed in the text.} 
\end{figure}

\section{Comparison to Previous Work}

The transport properties for charged particles moving through turbulent magnetic fields
was analyzed by Casse et al. (2002) using a method similar to that adopted in our work.
Specifically, these authors performed extensive numerical experiments using the
formalism developed by Giacalone \& Jokipii (1994) in order
to determine the pitch angle, scattering rate, and the parallel and perpendicular spatial diffusion
coefficients for a wide range of rigidities and turbulence levels.  Both parallel and perpendicular 
diffusion
coefficients are plotted versus rigidity $\rho = R_g k_{min} = 2 \pi / \bar\lambda_{max}$
for several different values of turbulence level 
\begin{equation} 
\eta = {\langle {\bf \delta B}^2 \rangle \over B_0^2 + \langle {\bf \delta B}^2 \rangle}\;.
\end{equation} 

We note that Casse et al. (2002) employed two different methods to construct their magnetic fields.
For $\eta = 1$ (which represents a purely turbulent field), these authors
adopted the same scheme presented in our work, and used a dynamic
range in wavelengths of $\lambda_{max}/\lambda_{min} = 10^4$.  
For all other cases, the magnetic field was constructed using a fast-Fourier
transform (FFT) algorithm to set up the magnetic field on a discrete grid in 
configuration space.  An interpolation scheme was then used to calculate the
field at any point in space.  For this latter method, $\lambda_{max}/\lambda_{min} = 128$
for most cases.

Our analysis extends the work of Casse et al. (2002) in two ways.  First, while these
authors focused exclusively on Kolmogorov diffusion, we also consider Bohm and Kraichnan diffusion.  
Second, we extend considerably the dynamic range of turbulence wavelengths, especially
for the case of a uniform field with underlying turbulence.  In addition, we focus our results
to the propagation of cosmic-rays in molecular cloud environments.  Nevertheless, sufficient
overlap exists for a direct comparison of a subset of our works.   
Specifically, experiments 19 - 25 listed in Table 1 (purely turbulent field) can be 
compared directly with the $\eta = 1$ data presented in Figure 4 of Casse et al. (2002).
In order to do so, we calculate the corresponding diffusion coefficients
\begin{equation}
{D\over R_g c} =  {\langle \Delta \bar x^2 \rangle \over 2 \tau_{max}}\;,
\end{equation}
where $\Delta \bar x$ is the particle 
displacement (from the origin) along the $x$ direction (although all directions are
equivalent) evaluated at time $\tau = \tau_{max}$.
We note that this method, while not exactly similar, is analogous to that adopted by Casse et al. (2002).
As shown in Figure 20, our results (open squares)
are in agreement with those of our predecessors (filled circles),
with the dotted line 
denoting the value of $\rho_{min} = 2\pi/\bar\lambda_{max}$ used in their calculations.
\begin{figure}
\figurenum{20}
{\centerline{\epsscale{0.90} \plotone{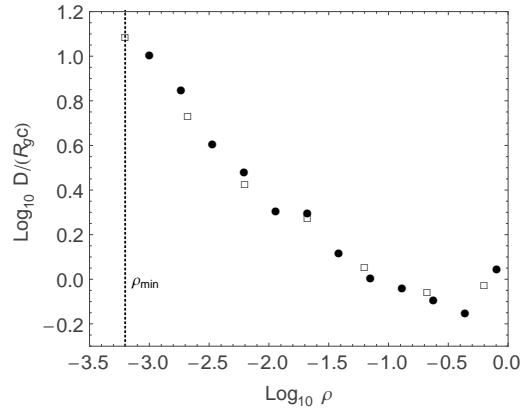} }}
\figcaption{Comparison of the diffusion coefficients calculated for experiments 19 - 25 listed in Table 1
(open squares) with the corresponding diffusion coefficients presented in Figure 4 of Casse et al. 2002
(filled circles).   The dotted vertical  line 
denotes the value of $\rho_{min} = 2\pi/\bar\lambda_{max}$ adopted by this earlier work for
the results shown here.} 
\end{figure}

We next compare our results from \S 6 for the case of a uniform field with underlying turbulence
to the $\eta = 0.46$ case presented in Figures 4 and 5 of Casse et al. (2002).  To do so,
we calculate both perpendicular and parallel diffusion coefficients
\begin{equation}
{D_{\perp}\over R_g c} =  {\langle \Delta \bar x^2 \rangle \over 2 \tau_{max}}\;,
\qquad
{D_{||}\over R_g c} =  {\langle \Delta \bar z^2 \rangle \over 2 \tau_{max}}\;,
\end{equation}
for experiments
15 - 21 in Table 2.
We compare our results (open squares and circles) to those of our
predecessors (filled squares and circles) in Figure 21.  The dotted line
denotes the value of  $\rho_{min} = 2\pi/\bar\lambda_{max}$ used by Casse
et al. (2002) for this case.   As expected, the results are in good agreement
for $\rho > \rho_{min}$, but deviate for lower values of rigidity, further  illustrating 
our conclusion from \S 5 that particle diffusion is not sensitive to the 
value of $\lambda_{min}$ so long as $R_g > \lambda_{min}$.
\begin{figure}
\figurenum{21}
{\centerline{\epsscale{0.90} \plotone{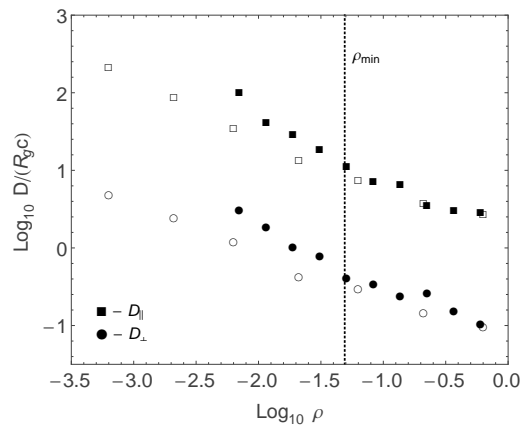} }}
\figcaption{Comparison of the parallel (open squares) and perpendicular (open circles) 
diffusion coefficients calculated 
for experiments 15 - 21 listed in Table 2
with the corresponding diffusion coefficients ($\eta = 0.46$) 
presented in Figures 4 and 5 of Casse et al. 2002
(filled squares and circles).   The dotted vertical  line 
denotes the value of $\rho_{min} = 2\pi/\bar\lambda_{max}$ adopted by this earlier work for
the results shown here.} 
\end{figure}

\section{Application to Cosmic-ray Diffusion in Molecular Clouds}

One of the original motivations for this calculation was to
determine what kind of injection profile would be required in order to correctly interpret the
apparent correlation between the diffuse $\gamma$-ray emissivity and the distribution of molecular
gas in the interstellar medium. Such a correlation between $\gamma$-ray intensity maps and the
large-scale features of the diffuse gas was first noted in observations ($E_{\gamma} \geq$ 100 MeV) 
with the SAS-2 and COS B satellite telescopes, combined with radio data that reveal the column 
density of interstellar hydrogen. Later observations associated at least
ten EGRET sources with SNRs expanding into MCs (Esposito et al. 1996;
Combi et al. 1998, 2001; Torres et al. 2003). 
More recently---and more spectacularly---a strong correlation
between TeV emission and the molecular gas distribution at the Galactic center was demonstrated
by HESS (Aharonian et al. 2006a;  Wommer et al. 2008). These data lend support to the idea that 
the low latitude $\gamma$-ray emission is mainly due to the decay of neutral pions produced by the 
scattering of cosmic rays with protons in the ambient medium rather than from bremsstrahlung or 
inverse Compton (IC) scattering.

In their assessment of this effect, Aharonian and Atoyan (1996) argued that the principal region
of interest for the $\pi^0$-decay $\gamma$-ray emission ought to lie within an $R \leq$ 100 pc region 
surrounding the cosmic-ray source. Within this distance of a ``typical" particle accelerator, a total energy 
output of $W_p \sim 10^{50}$ erg translates into a mean particle energy density of $w_p = W_p/(4/3) \pi 
R^3 \approx 0.55(W_p / 10^{50}\;{\rm erg})(R/100\;{\rm pc})^{-3}$ eV/cm$^3$, which may significantly exceed 
the average level of the ``sea" of galactic cosmic rays with energy density $w_0 \approx$ 1 eV/cm$^3$. 
Therefore, in a 1$^{\circ}$--10$^{\circ}$ region around a cosmic-ray source (depending on the distance to 
the source), we should expect to see higher than average $\gamma$-ray emission. In addition, if the diffusive 
propagation of cosmic rays is energy-dependent, the resulting $\gamma$-ray spectrum will differ from the 
$\gamma$-ray spectrum produced by galactic cosmic rays (e.g., Fujita et al. 2009). 
Thus, the possibility of having several dense giant 
molecular clouds (GMCs) in close proximity to a particle accelerator will not only produce higher than 
average levels of $\gamma$-rays but may give the appearance that there are multiple {\it distinct} 
cosmic-ray sources or, due to the limited angular resolution of instruments like EGRET, an {\it extended} 
cosmic-ray source. Accurately predicting the spatial and temporal evolution of the $\gamma$-ray spectrum 
produced by a particle accelerator may therefore lead to the classification of tens of unidentified EGRET 
sources.

In order to apply our results from \S \S 5 and 6 to molecular cloud environments, 
we consider the ideal case of a single {\it impulsive} cosmic-ray source 
surrounded by a homogeneous molecular cloud of radius $R$.
While the value of $\lambda_{\rm max}$ is not known
for such environments, one would expect its value to
be constrained from below by the size of dense cores ($\sim 0.1$ pc)
and from above by the size of the actual cloud ($\sim 10 - 20$ pc).  
We therefore adopt the intermediary value of $\lambda_{\rm max} = 1$ pc
in our discussion (although we keep $\lambda_{max}$ in our scaled
equations below).   The energy range $10^{-4} \le \eps \le 0.1$ 
of our work (as shown in Figures 14 and 19) thus corresponds to a
true particle energy range of $1 \la E_p \la10^3$ TeV.  
In turn, since only $\sim 10\%$ of a relativistic protons' energy 
goes into the $\pi_0$ photon decay channel for $pp$ scattering 
(see, e.g., Fatuzzo et al. 2006), the corresponding energy
range of $\gamma$-rays resulting from the interaction of these CR's and the
ambient molecular cloud medium is $0.1 \la \eps_\gamma \la 10^2$ TeV,
which falls within the range observable by HESS. 

As shown by Aharonian and Atoyan (1996), the energy loss rate of protons
with energies needed to produce 
$\pi^0$-decay $\gamma$-rays is dominated by nuclear energy losses due to
$pp$ scattering with the ambient medium. The 
lifetime of the protons, $\tau_{pp}$, depends on the pp-scattering cross-section, $\sigma_{pp}$, and 
the inelasticity parameter, $\kappa$. Over a broad range of proton energies, neither of these quantities 
significantly varies so the usual method is to adopt the constant average values $\sigma_{pp} \approx$ 
40 mb and $\kappa \approx$ 0.45 (see, e.g., Markoff et al. 1997). That being the case, the proton 
lifetime becomes independent of proton energy:
\begin{equation}
\tau_{pp} = (n c \kappa \sigma_{pp})^{-1} \approx 3 \times 10^5 \;{\rm yr}\,
\left({n_{H_2}\over 100\;{\rm cm}^{-3}}\right)^{-1} \;,
\end{equation}
where $n$ is the number density of ambient protons (i.e., $n = 2 n_{H_2}$).

We compare this timescale to the particle escape time $\tau_e$, defined
here as the time it takes 
CR's to diffuse a distance $\langle r \rangle = R$ for
purely turbulent fields, and $\langle z_{rms} \rangle = R$
if an underlying uniform field threads the molecular cloud.   
For  the intermediary case of Kraichnan diffusion, 
Equations 21 -- 23 can be combined to yield the expression
\begin{eqnarray}
\tau_{e; {\rm turb}} \approx 4\times 10^5\,{\rm yrs}
\,\left({R\over 20\,{\rm pc}}\right)^2\, \qquad \nonumber\\
\left({E_p\over 1\,{\rm TeV}}\right)^{-1/2}
\left({\lambda_{\rm max}\over 1 \,{\rm pc}}\right)^{-1/2}\,\left({B_0\over 10 \,\mu {\rm G}}\right)^{1/2}\,.
\end{eqnarray}
Likewise, Equations 21, 23 and 25 can be combined to
yield the expression
\begin{eqnarray}
\tau_{e; {\rm unif}} \approx 10^5\,{\rm yrs}
\,\left({R\over 20\,{\rm pc}}\right)^2\, \qquad \nonumber\\
\left({E_p\over 1\,{\rm TeV}}\right)^{-1/2}
\left({\lambda_{\rm max}\over 1 \,{\rm pc}}\right)^{-1/2}\,\left({B_0\over 10 \,\mu {\rm G}}\right)^{1/2}\,.
\end{eqnarray}
As suggested by Figure 1,  injected particles will therefore gain
a modest energy of $\sim 1$ TeV due to acceleration from
the turbulent electric fields before they escape.  As such, the fits
to the data presented in Figures 14 and 19 (as summarized by
Equations 22 -- 25) cannot be extrapolated to lower energies
for molecular cloud environments
(since $\eps = 10^{-4}$ represents a particle energy
of 0.92 TeV under the assumed conditions).  

Given the similarity between $\tau_{pp}$ and $\tau_{e}$,
a significant fraction of $> $ TeV CR's will likely undergo
$pp$ scattering before escaping from the molecular cloud 
environment.  As this fraction decreases with increasing
energy, the resulting $\gamma$-ray spectrum will be
softer  than that of the injected 
particle spectrum (e.g.,  Fujita et al. 2009).
In addition, if magnetic fields in molecular clouds are 
purely turbulent, then the break in the $\langle r \rangle$ -- $\eps$
data shown in Figure 14
at $\eps\sim 0.005$ -- which for the assumed conditions
corresponds to a value of $E_p \sim 50$ TeV -- would likely
produce a break in an observed $\gamma$-ray spectrum 
at around $\eps_\gamma\sim 5$ TeV.  Such a break would
not be observed if molecular clouds are threaded by an
underlying uniform magnetic field (see, e.g., Figure 19).  

The total $\gamma$-ray luminosity expected from our
assumed molecular 
cloud with a singe injection source $W_p$ is
independent of the escape time, as can be seen through the
simple estimate
\begin{eqnarray}
L_\gamma \approx f \left({\tau_e\over\tau_{pp}}\right)\, 
\left({W_p\over\tau_e}\right)
=  10^{36} \,{\rm erg}\,{\rm s}^{-1} \nonumber \\
\left({f \over 0.1}\right)\,\left({W_p\over 10^{50}\,{\rm erg}}\right)
\left({\tau_{pp}\over 3\times 10^5 {\rm yr}}\right)^{-1}\;,
\end{eqnarray}
where $f$ takes into account  that  only $\sim 10\%$ of the 
relativistic protons' energy goes into the $\pi_0$ photon decay channel.

Interestingly, this value is in reasonable agreement with the $\approx 10^{35}$
erg s$^{-1}$  luminosities in the 0.1 - 100 GeV band inferred for four SNR's 
interacting with molecular clouds (G349.7+0.2; CTB 37A; 3C 391; G8.7-0.1)
observed by the Large Area Telescope
on board the {\it Fermi Gamma-ray Space Telescope} (Castro \& Slane 2010).
Two of these SNRs (CTB 37A and G8.7-0.1) are also possible counterparts
to HESS  sources with implied luminosities in the 0.2 - 10 TeV
band of $\approx 5\times 10^{34}$ ergs s$^{-1}$ and $\approx 2\times10^{35}$ erg s$^{-1}$,
respectively, and three additional HESS sources coincident with SNRS G338.3-0.0,
G12.82-0.02 and W41
have implied 0.2 - 10 TeV luminosities of $\approx 2\times 10^{35}$ erg s$^{-1}$,
$3\times 10^{34}$ erg s$^{-1}$,
 and 
$4\times 10^{34}$ erg s$^{-1}$, respectively
(Aharonian et al. 2006b) .  Finally, we note that the 
above expected luminosity also falls within the range 
$1\times 10^{34}$--$4\times 10^{36}$ ergs s$^{-1}$ inferred from observations of 
the EGRET SNRs, although the energy range of this instrument 
only goes up to $\sim 30$ GeV.  

\section{Conclusion}  

We have investigated how high-energy CR's propagate through molecular 
cloud environments using a modified numerically based formalism 
developed by Giacalone \& Jokipii (1994) for the general study of cosmic-ray diffusion,
thereby providing a baseline analysis for two magnetic field configurations:  
1) a purely turbulent field; and 2) a uniform magnetic field with a strong turbulent 
component. We have focused most of our analysis on cases for which the particle 
gyration radius $R_g$ falls comfortably within the range of wavelengths shaping
the turbulence. For a purely turbulent field, the trajectory of a particle is fully
described by four dimensionless parameters.  However, we have found that the diffusion 
of an ensemble of particles through a turbulent field (characterized by the index
$\Gamma$) depends primarily on only one of these---the dimensionless scale length
$\bar\lambda_{\rm max}\equiv\lambda_{\rm max} / R_g$. For a uniform field with a 
turbulent component, CR diffusion depends on one additional dimensionless 
parameter---the ratio of turbulent field energy density to the uniform field 
energy density. 

Given the chaotic nature of particle motion through turbulent magnetic fields,
we performed a suite of statistical experiments as defined by the dimensionless 
parameters listed in Table 1 (a purely turbulent field) and Table 2 (a uniform field
plus a strong turbulent component). Specifically, we calculated the trajectory of $N_p$ 
particles injected randomly from the origin for a time $\tau_{max}$ for each experiment, 
with each particle sampling its own unique (and randomly selected) magnetic field 
structure. The resulting distributions of particle displacement along a given axis
were found to be well described by Gaussian profiles with the same mean and 
variance, thereby justifying our use of the mean of the particle displacements 
$\langle r\rangle$ as our output measure for characterizing the diffusion of 
particles through purely turbulent fields, and the rms values of the particle positions 
$x_{rms}$ and $z_{rms}$ as our output measures for a uniform 
magnetic field $B_0 \hat z$ with a strong turbulent component. We have found that 
after an initial time during which particles travel a distance $\sim \lambda_{\rm max}$, 
each of these output measures scales as $\sqrt{t}$, as expected for a diffusion process.  

The results of our analysis indicate that particle diffusion behaves differently
for gyro radii in the ranges $0.01 \lambda_{\rm max} \la R_g \la \lambda_{\rm max}$
and $R_g \la 0.01 \lambda_{\rm max}$.  Specifically, we have found that in the former
case, particles ``random walk" through the field, whereas for the latter, particles 
are strongly coupled to field lines and their motion is directly tied to the field 
line structure.  In turn, the distance over how far particles diffuse in purely 
turbulent fields as a function of energy exhibits a clear break at the point where 
the particle's gyration radius $R_g \approx 0.005\lambda_{\rm max}$.   

Comparing our results with those obtained in earlier works, we find good agreement
with previous results obtained using the same formalism (e.g., Casse et al. 2002).
In addition, our results are well-fit by the ``standard" scaling law
$R_{\rm diff} \propto E_p^{1-\Gamma/2}$ often invoked in the literature.
We provide simple scaling relations between mean diffusion lengths
and energy for both magnetic field profiles considered.
We note, however, that  these scaling-laws lead to a significant 
underestimation of the diffusion lengths
for the case of purely turbulent fields at energies 
$E_p \ga 0.005\, \lambda_{max} \,e\,B_0$.  In addition, the index
$1-\Gamma/2$ is 
not valid for the case of Bohm diffusion
($\Gamma = 1$) perpendicular to 
an underlying uniform magnetic field. 

The results of our work have important consequences 
for properly connecting $\gamma$-ray spectra associated with
molecular clouds to the underlying 
particle populations.  
We find that a significant fraction of $> $ TeV CR's will likely undergo
$pp$ scattering before diffusing out of a  molecular cloud 
environment.  As this fraction decreases with increasing
energy, the resulting $\gamma$-ray spectrum will be
softer  than that of the injected 
particle spectrum (e.g.,  Fujita et al. 2009).
In addition, if magnetic fields in molecular clouds are 
purely turbulent, then the break in the $\langle r \rangle$ -- $\eps$
dependence (as shown in Figure 14) is expected to 
produce a corresponding break in an observed $\gamma$-ray spectrum 
at around $\eps_\gamma\sim 5$ TeV.  Such a break would
not be observed if molecular clouds are threaded by an
underlying
uniform magnetic field.  

The work we have reported here has consequences for other types of high-energy
sources as well. For example, the compact object 1E~1740, embedded within a molecular 
cloud at the galactic center, produces a jet of (presumably) relativistic 
electrons and positrons (Misra \& Melia 1993) that eventually diffuse into the
surrounding medium. The diffuse radio inensity from this region provides some 
measure of the lepton injection rate, but it clearly also depends on the 
energy-dependent diffusion rate through the molecular gas. The results
reported here for proton diffusion cannot be directly generalized to the case
of positrons, but we anticipate seeing qualitative similarities between the two 
once we have completed the analogous positron simulations.

The galactic center hosts a complex array of diffuse emitters, in addition
to the TeV sources we have discussed in this paper. A proper analysis of the
underlying nonthermal particle population producing this emission should
therefore include observations at $\gamma$-ray (and even hard X-ray) energies,
in addition to the HESS data we have considered here (see, e.g., Belanger et 
al. 2004; Rockefeller et al. 2004). In future work, we will more closely 
examine the observational consequences of the different behavior of CR's 
above and below the break energy $E_b$, particularly as it impacts the 
diffuse broadband emission within $\sim$20 pcs of the supermassive black 
hole Sgr A*.

Of course, Sgr A* itself is apparently a significant accelerator of
relativistic electrons and protons (Liu et al. 2006), the latter diffusing
(Ballantyne et al. 2007) through the captured, accreting gas (Ruffert and
Melia 1994; Falcke et al. 1997) into the surrounding medium, possibly
producing the HESS point source coincident with the black hole. However,
attempts at reconciling this TeV emission with the longer wavelength
radiation produced closer to the center have been hampered by the uncertain
energy-dependence of this diffusion process. As we have discussed in this
paper, a detailed knowledge of the diffusion coefficient is essential for
meaningfully connecting the observed spectrum to the underlying nonthermal
particle population. We will be applying the conclusions reached here to
this important problem and will report the results elsewhere.

\acknowledgments

This work was supported by Xavier University through the Hauck Foundation, 
and by ONR grant N00014-09-C-0032 at the University of Arizona.  The authors would
like to thank the anonymous referee for several very useful comments that improved
the manuscript.






\begin{deluxetable}{rrrrrc}
\tablecolumns{6}
\tablewidth{0pc}
\tablecaption{Experiments for A Purely Turbulent Field}
\tablehead{
\colhead{Exp} & \colhead{$\Gamma$}   & \colhead{$\bar\lambda_{\rm min}$}    
& \colhead{$\bar\lambda_{\rm max}$} &
\colhead{$N_p$}  &  \colhead{$\tau_{max}/\bar\lambda_{\rm max}$}}
\startdata
1 & 3/2  & 3 & 300 & 1000 & $10^3$ \\
2 & 3/2  & 0.3 & 30 & 1000 &  $10^3$ \\
3 & 3/2  & 0.03 & 3 & 1000  & $10^3$ \\
4 & 3/2  & 0.1 & 10,000 & 200 & $10^2$\\
5 & 3/2  & 0.3 & 3,000 & 200 & $10^2$\\
6 & 3/2  & 0.1 & 1,000 & 200 & $10^2$\\
7 & 3/2  & 0.03 & 300 & 200 & $10^2$\\
8 & 3/2  & 0.01 & 100 & 200 & $10^2$\\
9 & 3/2  & 0.003 & 30 & 200 & $10^2$\\
10 & 3/2  & 0.001 & 10 & 200 & $10^2$\\
11 & 3/2  & 0.0003 & 3 & 200 & $10^2$\\
12 & 1  & 0.1 & 10,000 & 200 & $10^2$\\
13 & 1  & 0.3 & 3,000 & 200 & $10^2$\\
14 & 1  & 0.1 & 1,000 & 200 & $10^2$\\
15 & 1  & 0.03 & 300 & 200 & $10^2$\\
16 & 1  & 0.01 & 100 & 200 & $10^2$\\
17 & 1  & 0.003 & 30 & 200 & $10^2$\\
18 & 1  & 0.001 & 10 & 200 & $10^2$\\
19 & 5/3  & 0.1 & 10,000 & 200 & $10^2$\\
20 & 5/3  & 0.3 & 3,000 & 200 & $10^2$\\
21 & 5/3  & 0.1 & 1,000 & 200 & $10^2$\\
22 & 5/3  & 0.03 & 300 & 200 & $10^2$\\
23 & 5/3  & 0.01 & 100 & 200 & $10^2$\\
24 & 5/3  & 0.003 & 30 & 200 & $10^2$\\
25 & 5/3  & 0.001 & 10 & 200 & $10^2$\\

\enddata
\end{deluxetable}

\begin{deluxetable}{rrrrrrc}
\tablecolumns{7}
\tablewidth{0pc}
\tablecaption{Experiments for A Uniform Field Plus Strong Turbulence}
\tablehead{
\colhead{Exp} & \colhead{$\Gamma$}  
& \colhead{$\eta$} & \colhead{$\bar\lambda_{\rm min}$}    
& \colhead{$\bar\lambda_{\rm max}$} &
\colhead{$N_p$}  &  \colhead{$\tau_{max}/\bar\lambda_{\rm max}$}}
\startdata

1 & 3/2 &2 & 0.1 & 10,000 & 200 & $10^2$\\
2 & 3/2 &2 & 0.3 & 3,000 & 200 & $10^2$\\
3 & 3/2 &2 & 0.1 & 1,000 & 200 & $10^2$\\
4 & 3/2 &2 & 0.03 & 300 & 200 & $10^2$\\
5 & 3/2 &2 & 0.01 & 100 & 200 & $10^2$\\
6 & 3/2 &2 & 0.003 & 30 & 200 & $10^2$\\
7 & 3/2 &2 & 0.001 & 10 & 200 & $10^2$\\
8 & 1 &2 & 0.1 & 10,000 & 200 & $10^2$\\
9 & 1 &2 & 0.3 & 3,000 & 200 & $10^2$\\
10 & 1 &2 & 0.1 & 1,000 & 200 & $10^2$\\
11 & 1 &2 & 0.03 & 300 & 200 & $10^2$\\
12 & 1 &2 & 0.01 & 100 & 200 & $10^2$\\
13 & 1 &2 & 0.003 & 30 & 200 & $10^2$\\
14 & 1 &2 & 0.001 & 10 & 200 & $10^2$\\
15 & 5/3 &2 & 0.1 & 10,000 & 200 & $10^2$\\
16 & 5/3 &2 & 0.3 & 3,000 & 200 & $10^2$\\
17 & 5/3 &2 & 0.1 & 1,000 & 200 & $10^2$\\
18 & 5/3 &2 & 0.03 & 300 & 200 & $10^2$\\
19 & 5/3 &2 & 0.01 & 100 & 200 & $10^2$\\
20 & 5/3 &2 & 0.003 & 30 & 200 & $10^2$\\
21 & 5/3 &2 & 0.001 & 10 & 200 & $10^2$\\

\enddata
\end{deluxetable}

\begin{deluxetable}{crrr}
\tablecolumns{4}
\tablewidth{0pc}
\tablecaption{Fitting parameters for Figure 14}
\tablehead{
\colhead{$\Gamma$}  &&
\colhead{$\Lambda$}   & \colhead{$\alpha$} }
\startdata
1     && 2.2 & 0.5 \\
3/2  && 0.56 & 0.25 \\
5/3   && 0.35 & 0.17 \\
\enddata
\end{deluxetable}

\begin{deluxetable}{ccrrrrr}
\tablecolumns{7}
\tablewidth{0pc}
\tablecaption{Fitting parameters for Figure 19}
\tablehead{
\colhead{$\Gamma$} &&\colhead{$\Lambda_x$}   & \colhead{$\alpha_x$}
&&\colhead{$\Lambda_z$}   & \colhead{$\alpha_z$} }
\startdata
$1$& & 0.22 &0.36 && 3.2 & 0.5   \\
$3/2$& &0.22&0.25 && 1.1 &0.25   \\
$5/3$& &0.14&0.17 && 0.89 & 0.17   \\
\enddata
\end{deluxetable}

\end{document}